\documentclass{elsarticle}

\usepackage{epsfig}
\usepackage{a4wide}
\usepackage{amsmath}
\usepackage{wasysym}
\usepackage{array}
\usepackage{bm}

\begin{document}

\vspace{\baselineskip}

\title{The soft function for color octet production at threshold}

    \author{M. Czakon and P. Fiedler}

    \address{
      Institute for Theoretical Particle Physics and Cosmology, 
      RWTH Aachen University, D-52056 Aachen, Germany
    }

\cortext[thanks]{Preprint number: TTK-13-24}

\begin{abstract}

  \noindent
  We evaluate the next-to-next-to-leading order soft function for the
  production of a massive color octet state at rest in the collision
  of two massless colored partons in either the fundamental or the
  adjoint representation. The main application of our result is the
  determination of the threshold expansion of the heavy-quark
  pair-production cross sections in the quark annihilation and gluon
  fusion channels. We discuss the factorization necessary for this
  purpose and explain the relationship between hard functions and
  virtual amplitudes.

\end{abstract}

\maketitle

%%%%%%%%%%%%%%%%%%%%%%%%%%%%%%%%%%%%%%%%%%%%%%%%%%%%%%%%%%%%%%%%%%%%%%%%%%%%%%%%

\section{Introduction}

\noindent 
The recently completed calculation of the next-to-next-to-leading
order (NNLO) corrections to the total hadronic top-quark
pair-production cross section \cite{Baernreuther:2012ws,
  Czakon:2012zr, Czakon:2012pz, Czakon:2013goa} has been preceded by
the derivation of the velocity enhanced terms in a threshold expansion
in Ref.~\cite{Beneke:2009ye}. The latter publication exploited the
advances in soft-gluon resummation \cite{Czakon:2009zw, Beneke:2009rj}
and dealt with the incorporation of potential effects, which were
proven to factorize in Ref.~\cite{Beneke:2010da}. After factorization
of the Born cross section, the threshold expansion is given for the
two leading channels, quark annihilation and gluon fusion, in terms of
inverse powers of the top-quark velocity, $\beta$, and its
logarithms. Unfortunately, the expansion formulae are missing
$\beta$-independent terms, which are more difficult to derive. These
terms are of some phenomenological interest, since they propagate
through resummation to higher orders \cite{Cacciari:2011hy,
  Beneke:2011mq, Beneke:2012wb}. In principle, they can be
obtained by expanding the fits to the numerical results provided in
Refs.~\cite{Baernreuther:2012ws, Czakon:2013goa}. Nevertheless, due to
the inherent lack of numerical precision in the strict threshold
region, this approach leads to large uncertainties. In particular,
Ref.~\cite{Czakon:2013goa} quotes a 50 \% uncertainty on the constant
in the gluon fusion case. It is thus interesting, whether these
constant terms can be obtained by other methods.

Reviewing the same problem at the next-to-leading order (NLO), we
notice that the $\beta$-indepen\-dent terms have only been obtained
after an exact analytic calculation of the total cross sections
\cite{Czakon:2008ii}, and their projections onto the singlet and octet
color configurations of the top-quark pair
\cite{Czakon:2008cx}. Currently, it is hardly conceivable to perform
similar analytic calculations at NNLO. However, the threshold
behavior of cross sections for heavy-flavor production is much better
understood and it seems that the necessary information can be inferred
from soft-gluon factorization. In this case, the effect of the
radiation is contained in color configuration dependent soft
functions, which are convoluted with hard functions representing the
purely virtual contributions. As long as one is only interested in
NNLO expansions of cross sections including $\beta$-independent terms,
an additional factorization of the potential effects is not
necessary.

The hard and soft functions are needed for color singlet and octet
configurations of the final state. Unfortunately, apart from the color
singlet soft function \cite{Belitsky:1998tc}, they are unknown beyond
NLO. At NLO, the soft function for the production of a massive color
octet state at  threshold has been evaluated in
Refs.~\cite{Czakon:2009zw, Beneke:2009rj,
  Idilbi:2009cc}. Interestingly, Ref.~\cite{Idilbi:2009cc}
demonstrates its application to the production of a fundamental
scalar. The purpose of this work is to evaluate the color octet soft
function at NNLO. The hard functions will be presented together with
the complete virtual corrections in a subsequent publication.

Beyond total cross sections, there are related developments aiming at
the derivation of expansions and resummations of differential
distributions in different kinematical regimes of top-quark
pair-production. Most recent results are to be found in
Refs.~\cite{Ferroglia:2012ku,Ferroglia:2012uy, Ferroglia:2013zwa,
  Ferroglia:2013awa, Zhu:2012ts, Li:2013mia}. They are extensions of
previous analyses from Refs.~\cite{Ahrens:2011mw, Kidonakis:2010dk}.

The paper is organized as follows. In the next section, we discuss the
factorization of cross sections in the threshold limit with
emphasis on hard functions. Subsequently, we define the
soft function and provide the details of our calculation, the results
for the bare soft function in $d$-dimensions, as well as  the
renormalized expression to be used in applications. Conclusions and
outlook close the main text, which is supplemented with two
appendices, one on Wilson lines and the other containing the anomalous
dimensions, which are needed for renormalization.

%%%%%%%%%%%%%%%%%%%%%%%%%%%%%%%%%%%%%%%%%%%%%%%%%%%%%%%%%%%%%%%%%%%%%%%%%%%%%%%%

\section{Cross section factorization and hard functions}

\noindent 
Consider the total hadronic cross section for the production of a
heavy quark-anti-quark pair accompanied by any number of gluons
and massless quarks. We will denote with $Q$ the invariant mass of the
final state at threshold, i.e.\ $Q = 2m$, where $m$ is the heavy-quark
mass. Notice that our considerations can also be applied to the production
of an elementary state, e.g.\ a color octet scalar. In this case,
there are no potential exchange effects in the final state and the
discussion is slightly simplified.

The cross section can be written as
\begin{equation}
\sigma_{h_1 h_2} = \sum_{ab}
\hat\sigma^0_{ab} \otimes \phi^0_{a/h_1} \otimes
\phi^0_{b/h_2} \; ,
\end{equation}
where $\hat\sigma^0_{ab}$ is the partonic cross section for the
initial state partons $a$ and $b$, while $\phi^0_{a/h_1}$ and
$\phi^0_{b/h_2}$ are the parton distribution functions (PDFs) for the
partons $a$ and $b$ inside the hadrons $h_1$ and $h_2$. The
superscript 0 underlines that the quantities are not collinearly
renormalized, while the symbol $\otimes$ denotes convolution in the
momentum of the partons. There are only two possible channels in the
Born approximation to the partonic cross section: quark-anti-quark
annihilation and gluon-gluon fusion.

We now assume that we are only interested in the production close to
threshold, where the total energy of any additional radiation is
strongly restricted from above, and the final state heavy quarks are
non-relativistic. This condition can be enforced at the
hadronic level by the available collider energy. Nevertheless, we will
not discuss its phenomenological relevance in realistic
situations. Let us first ignore any potential (e.g.\ Coulomb)
interactions between the heavy quarks, which are also enhanced at
threshold. The partonic matrix elements factorize in this soft
limit. Any radiation can be approximated by emissions from eikonal
lines, which can be described as Wilson lines at the operator level
(see \ref{sec:wilson} for the definition and properties of Wilson
lines, and section \ref{sec:soft} for their relation to eikonal
lines).

In the soft approximation, the matrix elements of the basic hard $2 \to 2$
production process without radiation are taken at threshold (potential
effects are ignored at this point), while the
partonic cross section for this process is only affected by radiation
through the phase space volume. Let us denote the four-momentum of the
radiation by $P^\mu = (\omega, \vec p)$, while the initial parton
momenta in the center-of-mass system by $p_1$ and $p_2$. The phase
space volume for the two particle state depends on 
\begin{equation}
(p_1+p_2-P)^2 = \hat s-2\sqrt{\hat s} \, \omega + (\omega^2-\vec p^2) 
\approx \hat s - 2 \sqrt{\hat s} \, \omega
\; ,
\end{equation}
where $\hat s = (p_1+p_2)^2$, and the last approximation amounts to
only keeping the leading behavior in the $\omega \rightarrow 0$
limit. If we now parameterize $\omega$ as
\begin{equation}\label{eq:z1}
\omega = \frac{Q}{2}(1-z) \; ,
\end{equation} 
then the volume of the phase space will be given by
\begin{equation}\label{eq:z2}
\hat s-\sqrt{\hat s} Q(1-z) \approx \hat s z \; .
\end{equation}

\begin{figure}[h]
\begin{center}
\includegraphics[width=90mm,angle=0]{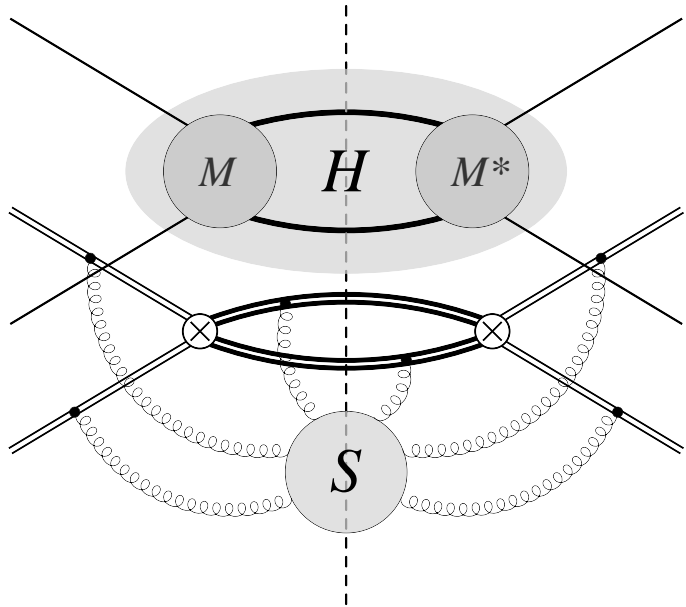}
\caption{\sf Schematic representation of soft
  factorization. $H$ stands for the hard function, whereas $S$ for the
  soft function. The double lines denote Wilson lines, whereas the
  $\otimes$ symbols stand for the insertion of the color structure of
  the hard matrix element. The sum over different color structures is
  suppressed.\label{fig:fac}}
\end{center}
\end{figure}

The factorization formula is
\begin{equation}\label{eq:fac1}
\hat \sigma^0 = \sum_{\alpha\beta} H^0_{\alpha\beta} \otimes
S^0_{\alpha\beta} \; ,
\end{equation}
where we have suppressed the parton indices, and the sum runs over
color structures $\alpha$ and $\beta$. $H^0_{\alpha\beta}$ are the hard
functions, i.e.\ cross sections for the $2 \to 2$ process, while
$S^0_{\alpha\beta}$ are the soft functions containing the effect of the
soft radiation from the Wilson lines. The convolution is performed in
the $z \in [0,1]$ variable defined by
Eqs.~(\ref{eq:z1},\ref{eq:z2}). The origin of the color structure
indices is best understood by inspecting a schematic representation of
the factorization given in Fig.~\ref{fig:fac}. The vertical dashed
line represents the unitarity cut. On the left hand side, we have the matrix
element, while on the right hand side its complex conjugate. The
factorization occurs at the matrix element level for each color
structure represented by the $\otimes$ symbol. The sum over color
configurations is thus coherent.

The choice of a basis for the color structures in the
factorization formula is crucial. Indeed, if the hard matrix
elements are decomposed into singlet and octet configurations of the
final state, then the out-going Wilson lines of the soft functions at
threshold, i.e.\ having the same velocity, can be combined into one as
shown in \ref{sec:wilson}. The  summation in Eq.~(\ref{eq:fac1})
becomes diagonal \cite{Beneke:2009rj}, since the color configurations
of the singlet and octet are orthogonal (for initial state gluons,
there are two octet configurations, symmetric and anti-symmetric -
they are also orthogonal by Bose symmetry). The final factorization
formula is now
\begin{equation}\label{eq:fac2}
\hat \sigma^0 = \sum_{\alpha} H^0_{\alpha} \otimes S^0_{\alpha} \; ,
\end{equation}
where the single color index in the hard and soft functions specifies
diagonal elements of both, and runs over singlet and octet configurations.

At this point, we have to take into account the effect of potential
interactions between the non-relativistic final state
quarks. Fortunately, it turns out that we do not have to make
substantial changes in our exposition \cite{Beneke:2010da}. As long as
the hard function contains $s$-wave effects only and is decomposed
into irreducible color representations of the final state,
non-relativistic effects factorize from the soft effects. This
factorization implies that the heavy-quark velocity must be set to
zero in the soft component, as we already assumed. On the other hand,
the hard amplitudes are to be expanded in the velocity, rather than
just evaluated at threshold. In principle, we could now further
factorize the potential effects in the hard functions as explained in
Ref.~\cite{Beneke:2010da}, but we shall not do that, since we are not
interested in their resummation, but rather in the fixed order
expansion of the cross sections at threshold. This is achieved by
formula Eq.~(\ref{eq:fac2}) after inclusion of the renormalization of
the PDFs in the soft limit in order to yield finite results.  A last
subtlety concerns the restriction to $s$-wave contributions in the
hard functions. Indeed, it is to be expected that higher partial waves
will occur in the velocity independent terms of their threshold
expansions beyond NLO. Nevertheless, they are not enhanced by soft
radiation at NNLO and, consequently, do not spoil the expansion
generated with Eq.~(\ref{eq:fac2}).

Although the partonic cross section on the left hand side of
Eq.~(\ref{eq:fac2}) suffers from initial state collinear divergences
only, factorization introduces additional divergences into the hard
and soft functions on the right hand side. We now consider the
renormalization of these infrared divergences of the hard functions. Since
the latter are given by total cross section contributions due to color
projected purely virtual amplitudes, the information we need is
contained in the singularities of the virtual amplitudes
themselves. It turns out that the complete divergence structure of a
UV renormalized amplitude $|M(\epsilon,\{\underline{p}\},
\{\underline{m}\})\rangle$ is encoded in the following equation 
\begin{equation}
{\bf
  Z}_M^{-1}(\epsilon,\{\underline{p}\},\{\underline{m}\},\mu)\,\,|M(\epsilon,\{\underline{p}\},\{\underline{m}\})\rangle\,=\,\text{finite}
\; ,
\end{equation} 
where the $\overline{\rm MS}$ renormalization constant ${\bf Z}_M$ is
a matrix in color space and has a non-trivial dependence on the 
kinematics $\{\underline{p}\}=\{p_1,...,p_n\}$, and by the same on the masses
$\{\underline{m}\}=\{m_1,...,m_n\}$ of the $n$ external partons. It
can be derived from the differential equation
\begin{equation}
\frac{d}{d\ln\mu}\,{\bf
  Z}_M(\epsilon,\{\underline{p}\},\{\underline{m}\},\mu)\,=\,-{\bf
  \Gamma}_M(\{\underline{p}\},\{\underline{m}\},\mu) \,
{\bf Z}_M(\epsilon,\{\underline{p}\},\{\underline{m}\},\mu) \; ,
\end{equation} 
where the color space matrix anomalous dimension is given by \cite{Ferroglia:2009ii}
\begin{equation}\label{eq:Gamma}
\begin{split}
{\bf \Gamma}_M(\{\underline{p}\},\{\underline{m}\},\mu)\,&=\,\sum\limits_{(i,j)}\frac{{\bf T}_i\cdot{\bf T}_j}{2}\,\gamma_{\text{cusp}}(\alpha_s)\,\ln\frac{\mu^2}{-s_{ij}}\,+\,\sum\limits_i \gamma^i(\alpha_s)\\
                                                       &-\,\sum\limits_{(I,J)}\frac{{\bf T}_I\cdot{\bf T}_J}{2}\,\gamma_{\text{cusp}}(\beta_{IJ},\alpha_s)\,+\,\sum\limits_I \gamma^I(\alpha_s)\,+\,\sum\limits_{I,j}{\bf T}_I\cdot{\bf T}_j\,\gamma_{\text{cusp}}(\alpha_s)\,\ln\frac{m_I\,\mu}{-s_{Ij}}\\
                                                       &+\,\sum\limits_{(I,J,K)}i\,f^{abc}\,{\bf T}_I^a\,{\bf T}_J^b\,{\bf T}_K^c\,F_1(\beta_{IJ},\beta_{JK},\beta_{KI})\\ 
                                                       &+\,\sum\limits_{(I,J)}\sum\limits_k\,i\,f^{abc}\,{\bf
  T}_I^a\,{\bf T}_J^b\,{\bf
  T}_k^c\,f_2\left(\beta_{IJ},\ln\frac{-\sigma_{Jk}\,v_J\cdot
  p_k}{-\sigma_{Ik}\,v_I\cdot p_k}\right)\,+\,\mathcal{O}(\alpha_s^3)
\; .
\end{split}
\end{equation} 

Its structure for massless partons has been determined already in
Ref.~\cite{Aybat:2006mz}. The structure of the massive case has been
studied in Refs.~\cite{Mitov:2009sv, Becher:2009kw}, with explicit
expressions for the second line given in Refs.~\cite{Becher:2009kw,
  Czakon:2009zw}. Finally, the third and fourth lines have been
determined in Ref.~\cite{Ferroglia:2009ii} (see also
\cite{Mitov:2010xw}).

The summations in Eq.~(\ref{eq:Gamma}) run over massless (indices $i, j,
k$) and massive (indices $I, J, K$) partons, with the notation $(i,j,...)$
denoting unordered tuples of different indices. The color operators
${\bf T}^a_i$ act on the color indices of the respective
partons.  If the particle is a gluon carrying a color index $c$, we
have $({\bf T}^a)_{bc}=-i\, f^{abc}$, assuming the result has been
projected on color index $b$. Similarly, for an outgoing
quark (or incoming anti-quark) the generator is $({\bf T}^a)_{bc} =
T^a_{bc}$, whereas for an incoming quark (or outgoing anti-quark) the
generator is $({\bf T}^a)_{bc} = -T^a_{cb}$. The kinematic dependence
is contained in $s_{ij} = 2\sigma_{ij} p_i \cdot p_j + i0^+$, where
the sign factor $\sigma_{ij} = +1$ if the momenta $p_i$ and $p_j$ are
both incoming or outgoing, and $\sigma_{ij} = -1$ otherwise. For
massive partons there is $p_I^2 = m_I^2$, $v_I = p_I/m_I$, and $\cosh
\beta_{IJ} = -s_{IJ}/2m_I m_J$. The cusp anomalous dimensions,
$\gamma_{\rm cusp}$, for the massless and massive cases, and the
functions $F_1,f_2$ can be found in Ref.~\cite{Ferroglia:2009ii} and
references therein.

It is interesting to note that the triple color correlations given in
the third and fourth lines of Eq.~(\ref{eq:Gamma}) cannot contribute
to the divergences of spin and color summed amplitudes at NNLO, as
long as the Born amplitudes do not contain complex couplings or
masses. This implies in particular that they will not contribute to
top-quark pair-production amplitudes, which was noticed for the quark
annihilation channel in Ref.~\cite{Czakon:2009zw} and for both channels in
Ref.~\cite{Ferroglia:2009ii}. In the general case, the argument is as
follows. First, notice that one can decompose the Born amplitude
treated as a vector in color and spin space in terms of color structures
\begin{equation}
| M^{(0)} \rangle = \sum_\alpha | M^{(0)}_\alpha \rangle \otimes | c_\alpha
\rangle \; ,
\end{equation}
where the vectors $| c_\alpha \rangle$ are made of $T^a_{bc}$ and $i
f^{abc}$ only. The amplitudes $|M^{(0)}_\alpha \rangle$ are
stripped of all color factors generated from QCD
vertices. Now, for $i,j,k$ all different (the indices make no
distinction this time between massive and massless partons), there is 
\begin{equation}\label{eq:antisymmetry}
\langle c_\alpha | \, if^{abc} {\bf T}^a_i {\bf T}^b_j {\bf T}^c_k \, |
c_\beta \rangle^* =
-\langle c_\beta | \, if^{abc} {\bf T}^a_i {\bf T}^b_j {\bf T}^c_k \,
| c_\alpha \rangle \; ,
\end{equation}
simply because the ${\bf T}^a_i$ operators are hermitian and commute
with each other as long as the parton indices are different, and
because the structure constants are real. On the other hand, both
sides of Eq.~(\ref{eq:antisymmetry}) are real, since they can be
evaluated with the Cvitanovi\'c algorithm
\cite{Cvitanovic:1976am} containing only real expressions. The color
matrix elements of the triple color correlator are thus anti-symmetric
in the color indices and we have
\begin{eqnarray}\label{eq:matrix}
\langle M^{(0)} | \, if^{abc} {\bf T}^a_i {\bf T}^b_j {\bf T}^c_k \, |
M^{(0)} \rangle &=& \sum_{\alpha\beta} \langle M^{(0)}_\alpha |
M^{(0)}_\beta \rangle \langle c_\alpha | \, if^{abc} {\bf T}^a_i {\bf
  T}^b_j {\bf T}^c_k \, | c_\beta \rangle \\ \nonumber 
&=& \frac{1}{2} \sum_{\alpha\beta} \left(\langle M^{(0)}_\alpha |
M^{(0)}_\beta \rangle - \langle M^{(0)}_\beta |
M^{(0)}_\alpha \rangle \right) \langle c_\alpha | \, if^{abc} {\bf T}^a_i {\bf
  T}^b_j {\bf T}^c_k \, | c_\beta \rangle
\; .
\end{eqnarray}
Since $\langle M^{(0)}_\alpha | M^{(0)}_\beta \rangle^* = \langle
M^{(0)}_\beta | M^{(0)}_\alpha \rangle$, the right hand side of
Eq.~(\ref{eq:matrix}) vanishes if $\langle M^{(0)}_\alpha |
M^{(0)}_\beta \rangle$ is real. Due to elementary spin summation
rules, this is the case if there are no complex parameters in the
Lagrangian (the case of complex parameters is generally of interest,
since we might want to describe unstable particles).

The argument presented above is, of course, also valid for diagonal matrix
elements between color projected Born amplitudes, which shows that the NNLO
renormalization constants for our hard functions can be determined from
the dipole correlations given in the first two lines of
Eq.~(\ref{eq:Gamma}). Since the initial partons are either in the
fundamental, or the adjoint representation, while the final
state may be in a singlet or octet configuration, we have to consider a
set of color vectors $| c^{\bm{R} \otimes \bm{\bar R} | {\bm{R'}}}
\rangle$, where $\bm{R} \in \{ \bm{3}, \bm{8} \}$ denotes the representation
of the initial partons, while $\bm{R'} \in \{ \bm{1}, \bm{8},
\bm{8_A}, \bm{8_S} \}$ that of the final state. The subscripts $\bm{S}$ and
$\bm{A}$ in the latter case stand for symmetric and anti-symmetric octets in
the case of an $\bm{8}\otimes\bm{8}$ initial configuration. The bare
hard functions for heavy flavor production are given by
\begin{equation}
H^0_{\bm{R} \otimes \bm{\bar R} | {\bm{R'}}} = {\cal N}
\int d \mbox{PS}_2
\sum_{\alpha\beta}
\frac{\langle \alpha | c^{\bm{R} \otimes \bm{\bar R} |
  {\bm{R'}}} \rangle \langle c^{\bm{R} \otimes \bm{\bar R} |
  {\bm{R'}}} | \beta \rangle}{\langle
  c^{\bm{R} \otimes \bm{\bar R} | {\bm{R'}}} | c^{\bm{R} \otimes
    \bm{\bar R} | {\bm{R'}}} \rangle}
 \langle M_\alpha | M_\beta \rangle
\; ,
\end{equation}
where the right hand side is expanded in the heavy-quark velocity,
$\beta$, up to and including terms of order $\beta$. ${\cal N}$
denotes the product of the flux, and color and spin average factors, while $| M_\alpha
\rangle$ are the purely virtual UV renormalized amplitudes for heavy
flavor production, where the initial state is specified by the color
representation $\bm{R}$. The expansion in $\beta$ in the definition of the
hard functions would directly correspond to the soft approximation of
taking the matrix element at threshold and keeping the exact phase
space, if there were no potential effects. Due to the latter, the
result contains inverse powers and logarithms of $\beta$.

The renormalization of the hard functions is now achieved with
\begin{equation}
H^0_{\bm{R} \otimes \bm{\bar R} | {\bm{R'}}} = Z^{\bm{R} \otimes
  \bm{\bar R} | {\bm{R'}}}_H(\mu/Q) \, H_{\bm{R} \otimes \bm{\bar R} |
  {\bm{R'}}} \; ,
\end{equation}
where the renormalization constant $Z_H^{\bm{R} \otimes \bm{\bar R} |
  {\bm{R'}}}$ satisfies the equation
\begin{equation}
\frac{d}{d \ln \mu} \, Z_H^{\bm{R} \otimes \bm{\bar R} |
  {\bm{R'}}}(\mu/Q) = -\Gamma_H^{\bm{R} \otimes \bm{\bar R} |
  {\bm{R'}}}(\mu/Q) \, Z_H^{\bm{R} \otimes \bm{\bar R} | {\bm{R'}}}(\mu/Q) \; ,
\end{equation}
with
\begin{equation}
\Gamma_H^{\bm{R} \otimes \bm{\bar R} |
  {\bm{R'}}}(\mu/Q) = \lim_{\beta \to 0} 2\,\Re\left( \langle c^{\bm{R}
  \otimes \bm{\bar R} | {\bm{R'}}} |\,{\bf
  \Gamma}_M\left(\beta,\cos\theta,\frac{\mu}{Q} \right)
| c^{\bm{R} \otimes \bm{\bar R} | {\bm{R'}}} \rangle\Big/
\langle
  c^{\bm{R} \otimes \bm{\bar R} | {\bm{R'}}} | c^{\bm{R} \otimes
    \bm{\bar R} | {\bm{R'}}} \rangle \right) \; .
\end{equation} 
Contrary to the hard function itself, the hard anomalous dimensions are
finite in the $\beta \to 0$ limit, and do not depend on the
scattering angle $\theta$. They can be found in
\ref{sec:AnomalousDimensions}.

We mentioned before that the same formalism can be applied to the
production of a fundamental object, e.g.\ a color octet
scalar. Clearly, the constant $Z_H^{\bm{R} \otimes \bm{\bar R} |
  {\bm{R'}}}$ will also be used for the renormalization of the soft
function, which does not depend on the nature of the final state. This
means that the latter does not play any role in the divergences of the
amplitude at threshold. Indeed, this is true for the real part of the
anomalous dimension ${\bf \Gamma}_M$ (the imaginary part contains
Coulomb phases). The final state anomalous dimension coefficients in
Eq.~(\ref{eq:Gamma}) for massive states have a purely soft origin. 

%%%%%%%%%%%%%%%%%%%%%%%%%%%%%%%%%%%%%%%%%%%%%%%%%%%%%%%%%%%%%%%%%%%%%%%%%%%%%%%%

\section{Soft function}
\label{sec:soft}

\noindent
We define the bare soft function for color octet production at rest as follows
\begin{eqnarray}\label{eq:SoftFunction}
S_{\bm{R} \otimes \bm{\bar R} | {\bm{R'}}}^0(\omega) &=& \frac{Q}{2}
\sum_{X} \delta(\omega - E_X) \\ \nonumber
&\times& \sum_{abc} \left| \sum_{a'b'c'}
C^{\bm{R} \otimes \bm{\bar R} | {\bm{R'}}}_{a'b'c'}
\langle X | T \left[
\Phi^{(\bm 8)}_{v,aa'}(+\infty,0)
\Phi^{(\bm R)}_{n,b'b}(0,-\infty)
\Phi^{(\bm{\bar R})}_{\bar n,c'c}(0,-\infty)
\right] | 0 \rangle\right|^2 \; ,
\end{eqnarray}
where the summation in the first line is taken over all possible states $X$
with energy $E_X$, and in the second line over color indices.
The Wilson line operators $\Phi^{(\bm R)}_{\beta,cd}(b,a)$
discussed in \ref{sec:wilson} have directions $n^\mu
= (1,\vec{\bm 0}^{(d-2)},1)$, $\bar{n}^\mu = (1,\vec{\bm 0}^{(d-2)},-1)$ for the
two incoming light-cone states, and $v^\mu = (1,\vec{\bm 0}^{(d-1)})$ for
the outgoing octet at rest.
Notice that time ordering is in fact only necessary for the product of
the two incoming lines. The notation $\bm{R} \otimes \bm{\bar R} | {\bm{R'}}$ specifies the
representation of the initial states as $\bm{R}$ and $\bm{\bar R}$,
and the restriction to the hard scattering color structure
corresponding to the irreducible component $\bm{R'}$ of the tensor
product $\bm{R} \otimes \bm{\bar R}$. We consider three cases
\begin{equation}\label{eq:ColorStructures}
C^{\bm{3} \otimes \bm{\bar{3}} | {\bm{8}}}_{abc} =
\sqrt{\frac{2}{N_c^2-1}} T^a_{cb} \; , \;\;
C^{\bm{8} \otimes \bm{8} | {\bm{8_A}}}_{abc} =
\sqrt{\frac{1}{N_c(N_c^2-1)}} i f^{abc} \; , \;\;
C^{\bm{8} \otimes \bm{8} | {\bm{8_S}}}_{abc} =
\sqrt{\frac{N_c}{(N_c^2-1)(N_c^2-4)}} d^{abc} \; ,
\end{equation}
where $N_c = 3$, and the symmetric tensor $d^{abc}$ is defined through
$\mbox{Tr} \; T^a T^b T^c = 1/4(d^{abc}+i f^{abc})$. The invariance of
the color structures together with the gauge transformation properties
of the Wilson lines, Eq.~(\ref{eq:covariance}), assure the gauge
invariance of the soft function.

Up to NNLO, the singularities in $\epsilon$ of $S_{\bm{R} \otimes
  \bm{\bar R} | {\bm{R'}}}^0$ are known to only depend on the Casimir
invariants of the representations of the Wilson lines
\cite{Czakon:2009zw, Beneke:2009rj}. We have checked explicitly that
this is valid for the exact $\epsilon$ dependence as well. Since the
out-going massive Wilson line is always in the octet representation,
the result depends on $\bm{R} \otimes \bm{\bar R} | {\bm{R'}}$ only
through $C_R$. Therefore, we will drop the subscript in $S_{\bm{R}
  \otimes \bm{\bar R} | {\bm{R'}}}^0$ in all the subsequent formulae.

The prefactor $Q/2$ in Eq.~(\ref{eq:SoftFunction}) and the normalization of the
color structures in Eq.~(\ref{eq:ColorStructures}) guarantee that the
soft function reduces to $\delta(1-z)$ at leading order. Dimensional
analysis allows to write the following perturbative expansion
\begin{equation}\label{eq:S0}
S^{0}(z) = \delta(1-z)+\frac{1}{1-z}\,\sum\limits_{n=1}^\infty
\left(\frac{Z_{\alpha_s}\,\alpha_s}{\pi}\right)^{n}\,\left(\frac{\mu}{Q\,(1-z)}\right)^{2
  n\epsilon}\,s^{(n)} \; ,
\end{equation} 
where the coefficients $s^{(n)}$ depend on $\epsilon$, color (through
$C_A$ and $C_R$) and the number of light quark flavors $n_f$ (as
usual, through $T_F n_f$).

We now apply a Mellin transform
\begin{equation}
\begin{split}
S^{0}(N)&= 1+\,\sum\limits_{n=1}^\infty \left(\frac{Z_{\alpha_s}\,\alpha_s}{\pi}\right)^{n}\,\left(\frac{\mu}{Q}\right)^{2 n\epsilon}\,\left[\int\limits_0^1 z^{\tilde{N}-1}\,(1-z)^{-1-2 n \epsilon}\,dz\right]\,s^{(n)}\\
                &= 1+\,\sum\limits_{n=1}^\infty \left(\frac{Z_{\alpha_s}\,\alpha_s}{\pi}\right)^{n}\,\left(\frac{\mu}{Q}\right)^{2 n\epsilon}\,\frac{\Gamma(\tilde{N})\,\Gamma(-2 n \epsilon)}{\Gamma(\tilde{N}-2 n \epsilon)}\,s^{(n)}\\
                &= 1+\,\sum\limits_{n=1}^\infty
\left(\frac{Z_{\alpha_s}\,\alpha_s}{\pi}\right)^{n}
\,e^{2n\epsilon(L-\gamma_E)}\,\Gamma(-2 n
\epsilon)\,s^{(n)}\,+\,\mathcal{O}\left(\frac{1}{N}\right) \; ,
\end{split}
\end{equation} 
where $\tilde{N} = N e^{-\gamma_{{E}}}$ and $L = \ln(\mu  N/Q)$. We
only keep the leading behavior in $1/N$, since the limit $N
\rightarrow \infty$ in Mellin space corresponds to the soft limit $z
\rightarrow 1$ . The last line of the above
equation demonstrates that the soft function depends on $L$ rather
than separately on $\mu/Q$ and $N$. The finiteness of the partonic
cross section implies that the renormalized soft function is
given by
\begin{equation}\label{eq:SR}
S(L) = Z_H(\mu/Q) Z_\phi^2(N) S^{0}(L) \; ,
\end{equation}
where we have suppressed the dependence on the initial state parton in
both $Z_H$ and $Z_\phi$, the latter being the renormalization constant
of the parton distribution function. $Z_\phi$ satisfies the equation
\begin{equation}
\frac{dZ_\phi(N)}{d\ln\mu^2} = -P(N)Z_\phi(N) \; ,
\end{equation}
where $P(N)$ is the soft expansion of the Altarelli--Parisi splitting
kernel in Mellin space, which can be found in
\ref{sec:AnomalousDimensions}. In consequence, the soft function obeys
the Renormalization Group Equation (RGE)
\begin{equation}\label{eq:RGE}
\frac{d S(L)}{dL} = -\left(\Gamma_H(\mu/Q) + 4 P \left(N\right)\right) S(L) \; ,
\end{equation}
which shows that the logarithmic dependence on $\mu/Q$ in $\Gamma_H$
and on $N$ in $P$ must combine into a dependence on their product as
given in $L$. This is a demonstration of the well known fact that the
singular part of the soft limit of the splitting kernels is given by
the same soft anomalous dimension, which governs the soft-collinear
singularities of the virtual amplitudes.

The RGE Eq.~(\ref{eq:RGE}) can be used to resum large logarithms of
$N$. For this purpose, it is sufficient to evolve from the scale $\mu
= \mu_0/N$, where $L = \ln(\mu_0/Q)$ is small, to the actual scale $\mu =
\mu_0$. We are, however, concerned with the fixed order perturbative
expansion, and will now present the calculation and results at NNLO.

\subsection{Calculation up to ${\cal O}(\alpha_s^2)$}

The ${\cal O}(\alpha_s)$ contribution to the soft function for color
octet production at threshold has been obtained in
\cite{Idilbi:2009cc, Czakon:2009zw}, and for general
representations of the three Wilson lines in
\cite{Beneke:2009rj}. With the conventions of Eq.~(\ref{eq:S0}), the
result reads
\begin{equation}
s^{(1)} =
-e^{\gamma_{{E}}\epsilon}\frac{\Gamma(1-\epsilon)}{\Gamma(1-2\epsilon)}\left(C_A\,\frac{1}{1-2
  \epsilon}+\,C_R\,\frac{2}{\epsilon}\right) \; .
\end{equation} 

The ${\cal O}(\alpha_s^2)$ contribution is conveniently evaluated in
the momentum representation, where Wilson lines become eikonal
lines. Each emission of a gluon with momentum $q_1$ from a hard 
parton with momentum $p$ contributes a factor
\begin{equation}
i g_s^0 \frac{i p^\mu}{\pm p \cdot q + i\epsilon} \, {\bf T}^{(\bm R)
  \, a} \; ,
\end{equation}
where $q$ is the sum of $q_1$ and the momenta of the gluons emitted
before for an in-going line (minus sign in the denominator), or after
for an out-going line (plus sign in the denominator). The phase space
integrations are performed with
\begin{equation}
\begin{split}
d\mbox{PS}_{{1}} &= \frac{Q}{2} \int
\frac{d\Omega_{{d-1}}\,dE}{(2\pi)^{d-1}}\,\frac{E^{d-3}}{2}\,\delta(\omega-E)
\; , \\ 
d\mbox{PS}_{{2}} &= \frac{Q}{2} \int
\frac{d\Omega_{{d-1}}^{(1)}\,dE_{{1}}}{(2\pi)^{d-1}}\frac{d\Omega_{{d-1}}^{(2)}\,dE_{{2}}}{(2\pi)^{d-1}}\,\frac{E_{{1}}^{d-3}}{2}\,\frac{E_{{2}}^{d-3}}{2}\,\delta(\omega-E_{{1}}-E_{{2}})
\; ,
\end{split}
\end{equation}
for the real-virtual (one-loop corrections to single gluon emission),
and double-real (double-gluon or quark-anti-quark emission) cases
respectively. For a gluon pair in the final state, an additional
factor of $1/2$ has to be included. 

Our result for the bare ${\cal O}(\alpha_s^2)$ contribution in
$d$-dimensions is presented in the form of four contributions
\begin{equation}
s^{(2)} = s_{{\Square}}^{(2)} + s_{{\bigtriangleup}}^{(2)} +
s_{{\Circle}}^{(2)} + s_{{\CIRCLE}}^{(2)} \; .
\end{equation} 
The first three of them correspond to double-real radiation, whereas
the last one to the real-virtual corrections. The two cases are
discussed separately.

\subsection{Double-real corrections}

\begin{figure}[h]
\begin{center}
\begin{tabular}{|m{1cm} m{3cm} m{3cm} m{3cm}|}
\hline
&&&\\
&\includegraphics[width=29mm,angle=0]{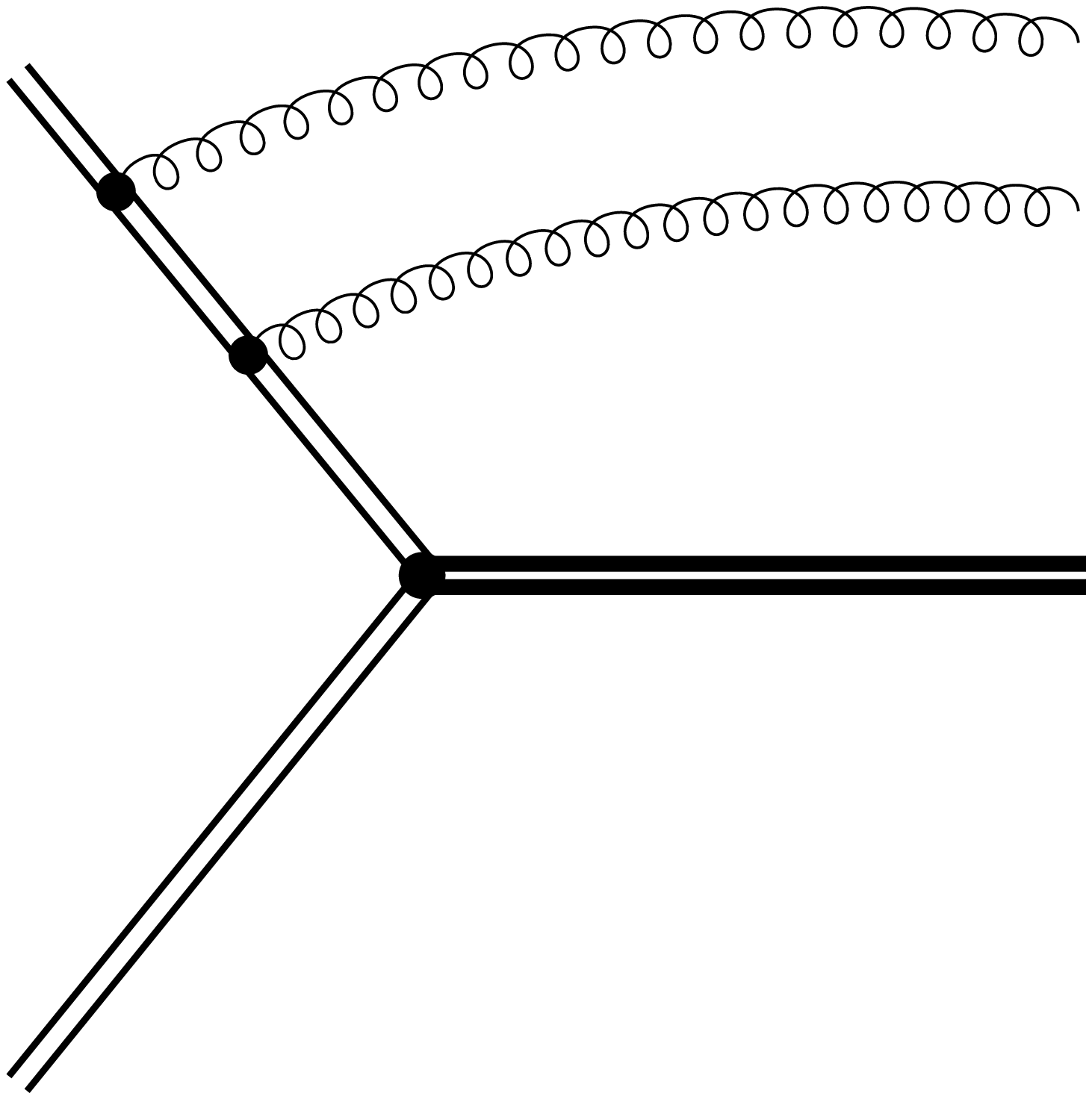} &\includegraphics[width=29mm,angle=0]{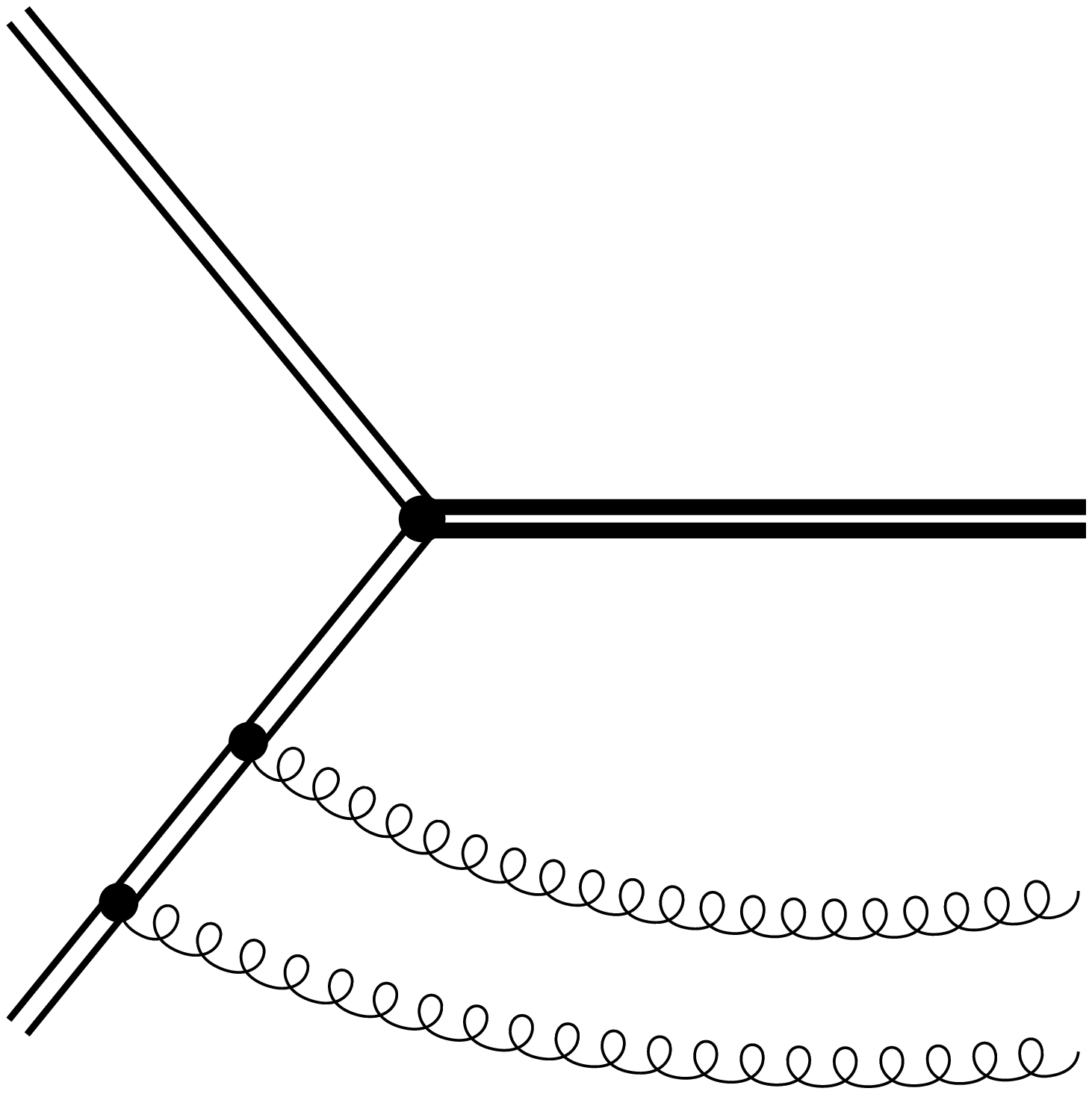}&\includegraphics[width=29mm,angle=0]{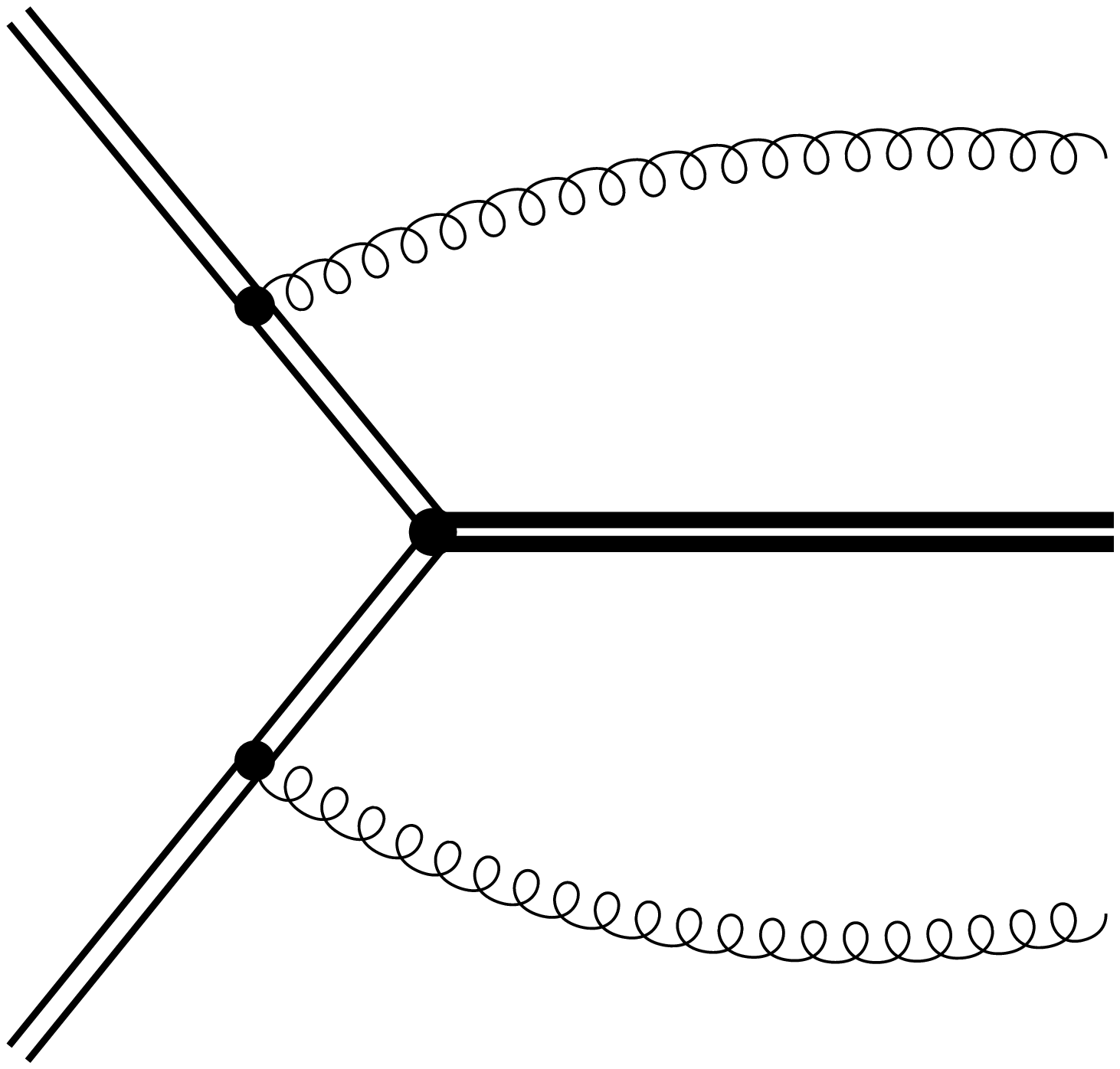}\\
(A)&&&\\
&\includegraphics[width=29mm,angle=0]{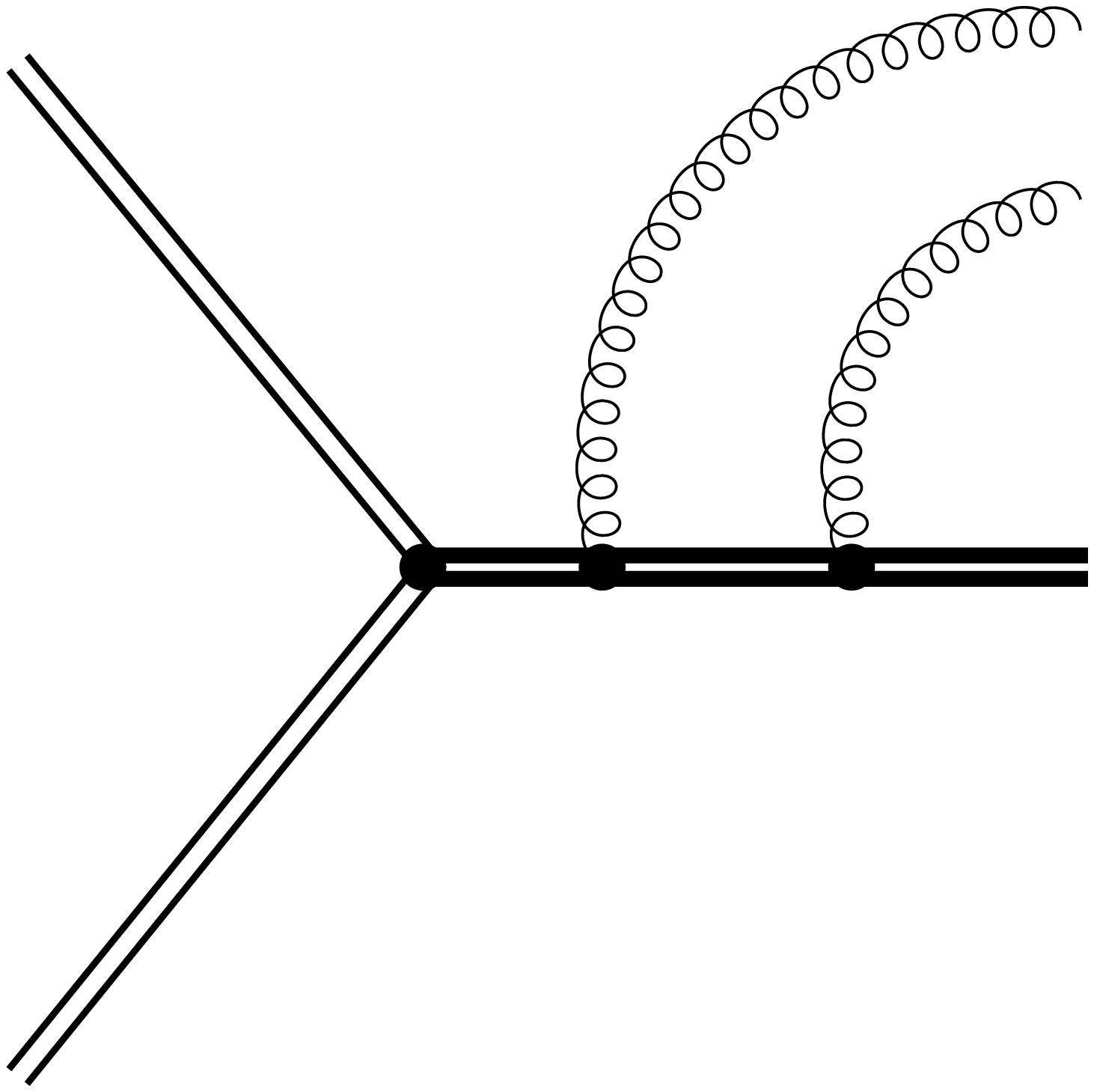} &\includegraphics[width=29mm,angle=0]{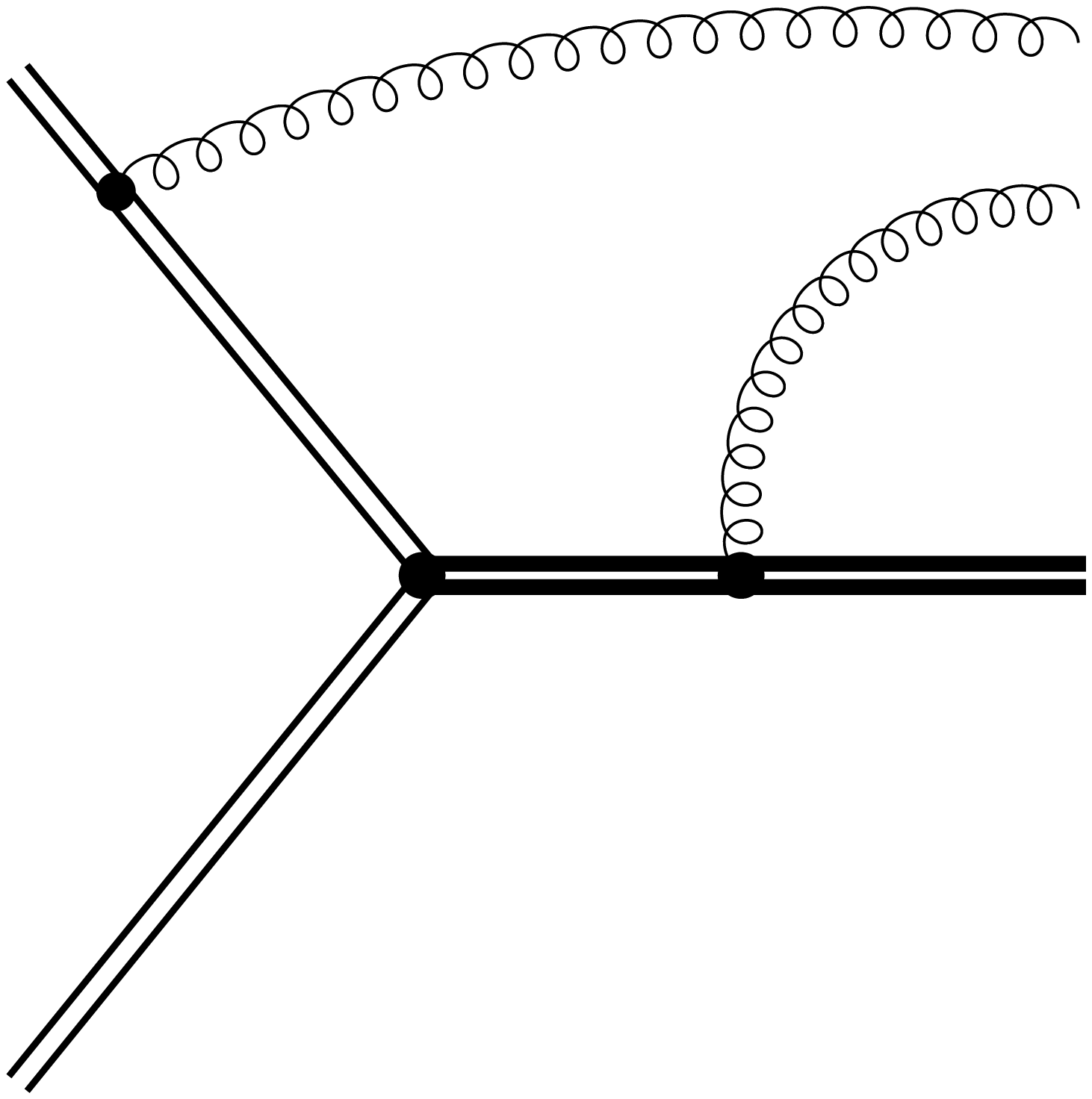}&\includegraphics[width=29mm,angle=0]{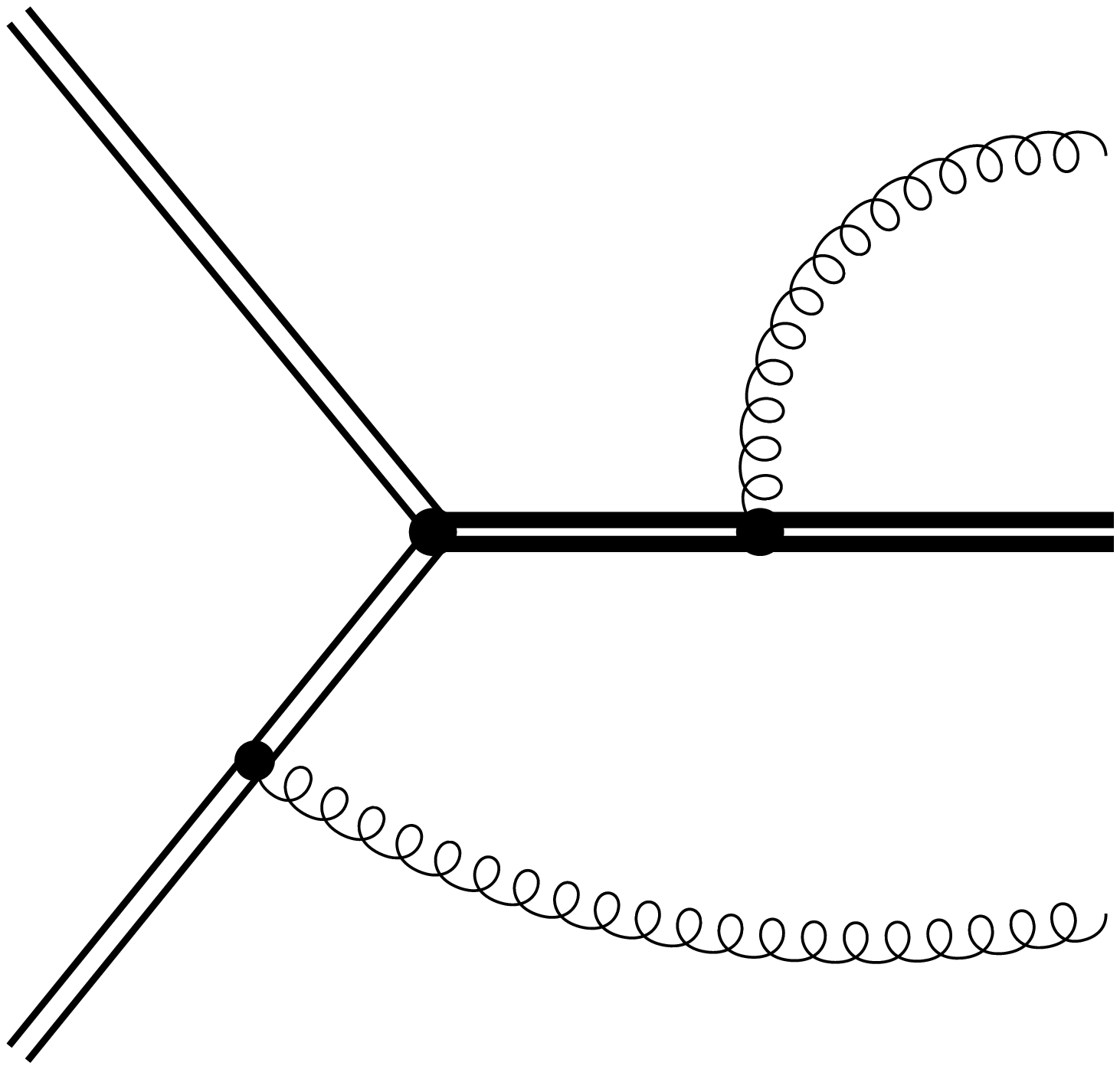}\\
\hline
&&&\\
(B)&\includegraphics[width=29mm,angle=0]{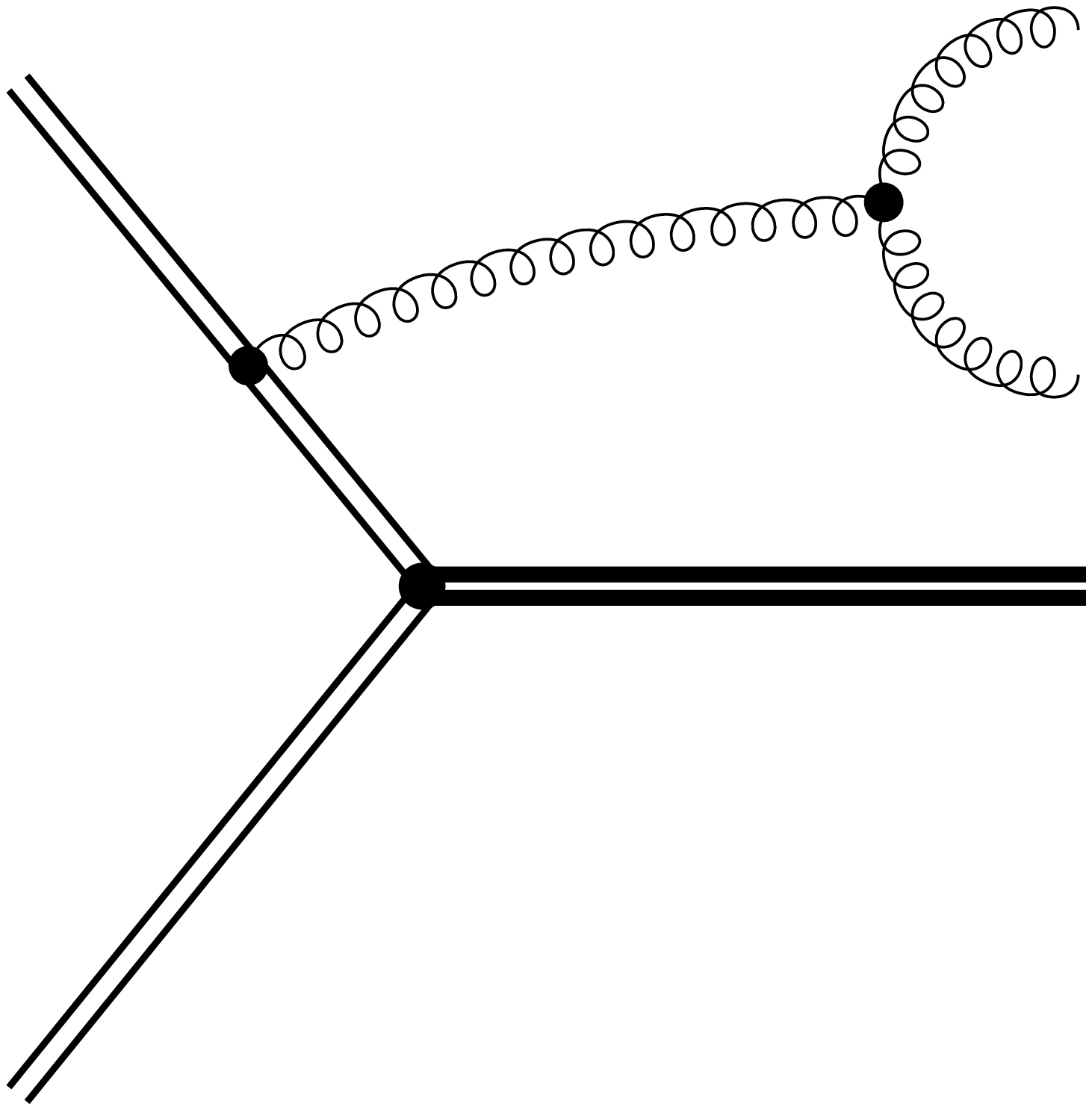} &\includegraphics[width=29mm,angle=0]{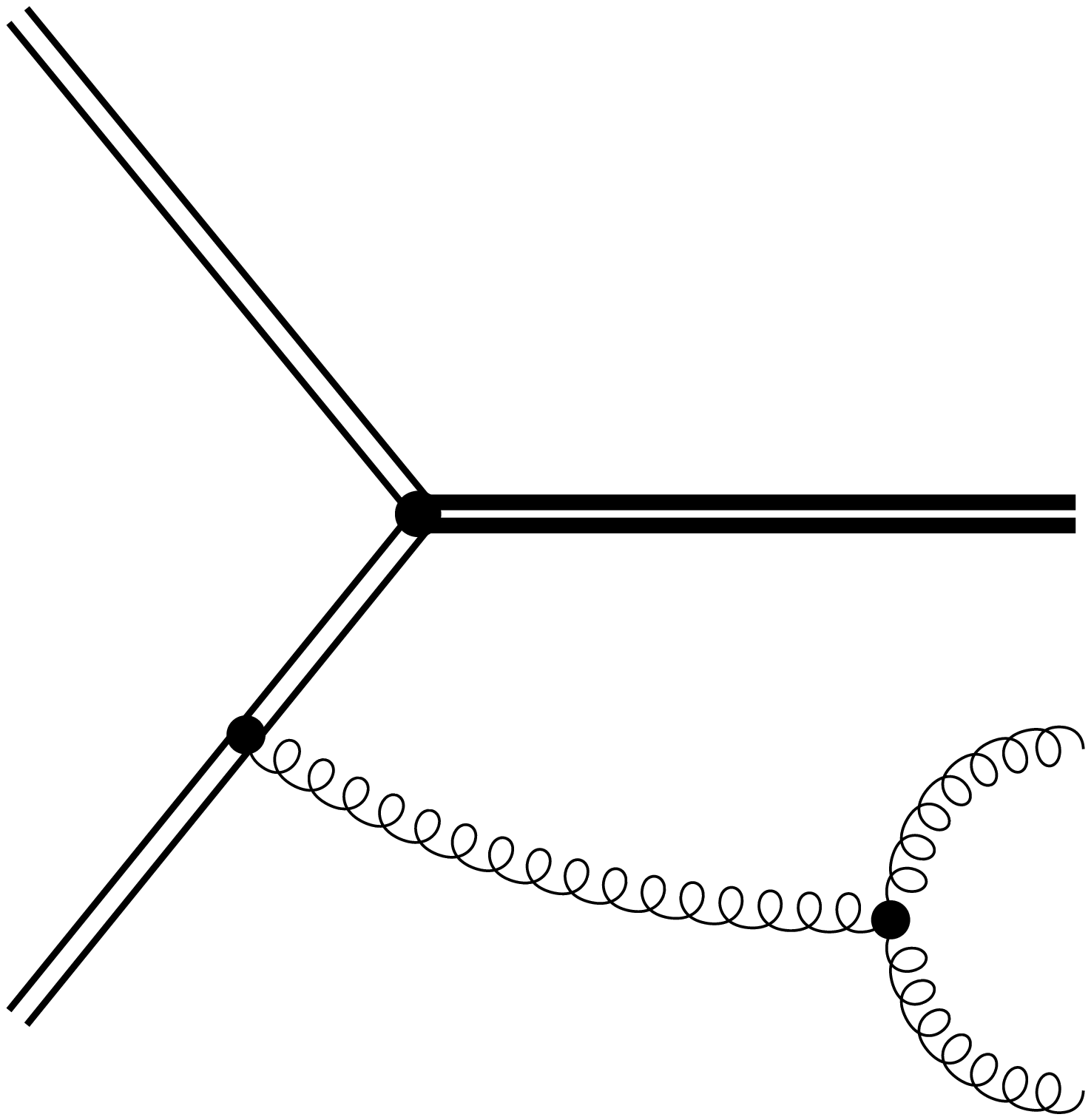}&\includegraphics[width=29mm,angle=0]{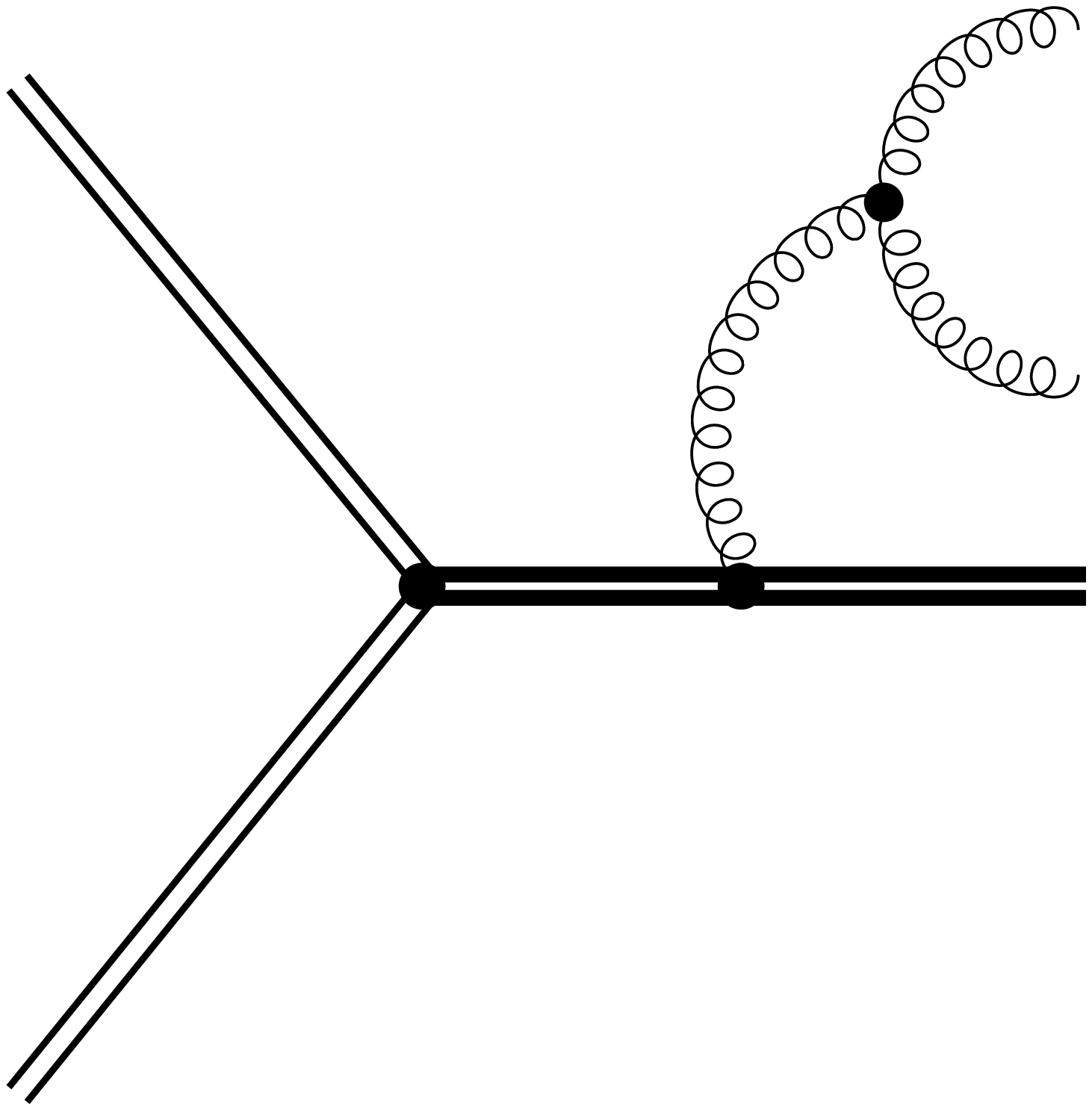}\\
\hline
&&&\\
(C)&\includegraphics[width=29mm,angle=0]{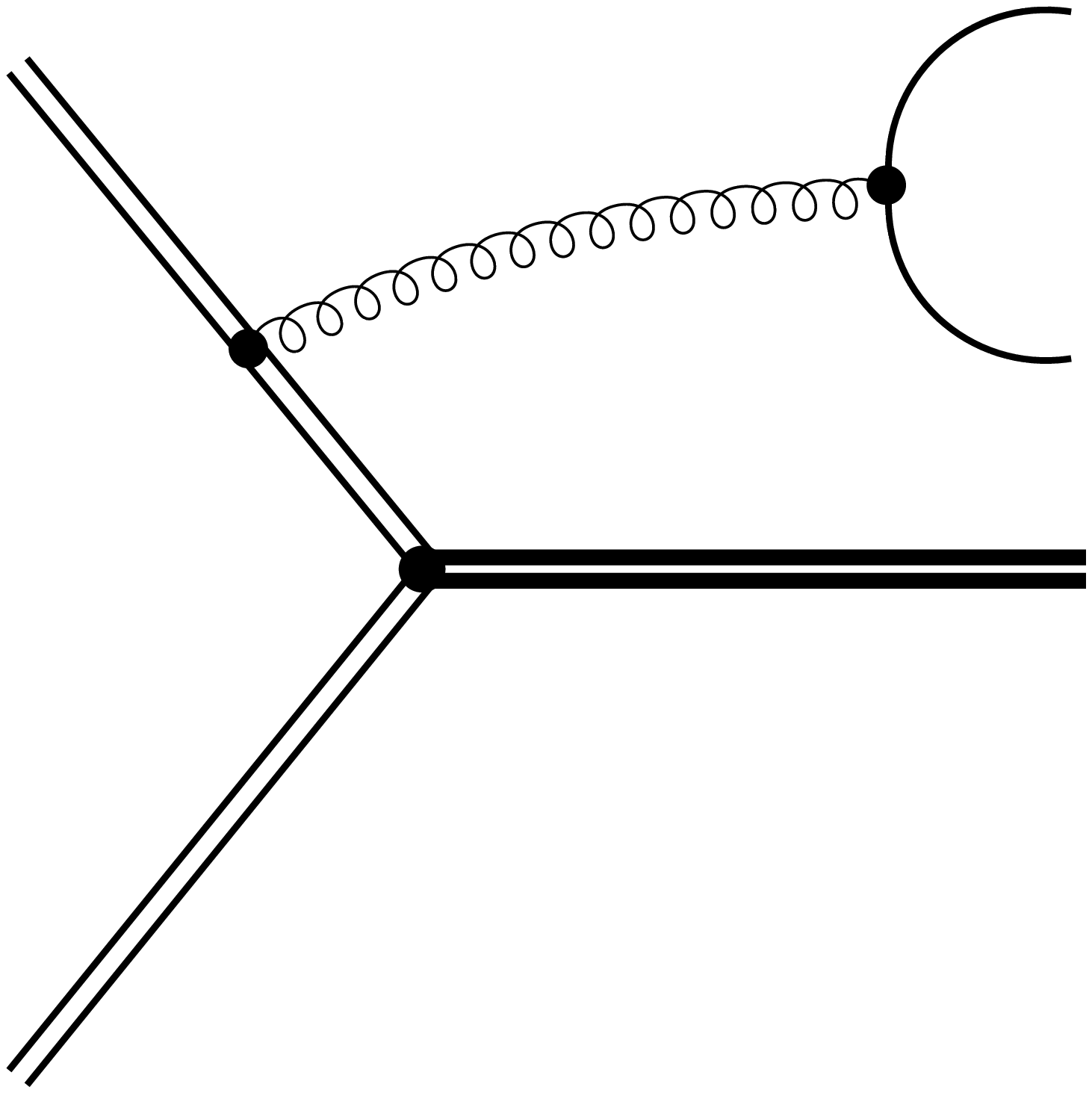} &\includegraphics[width=29mm,angle=0]{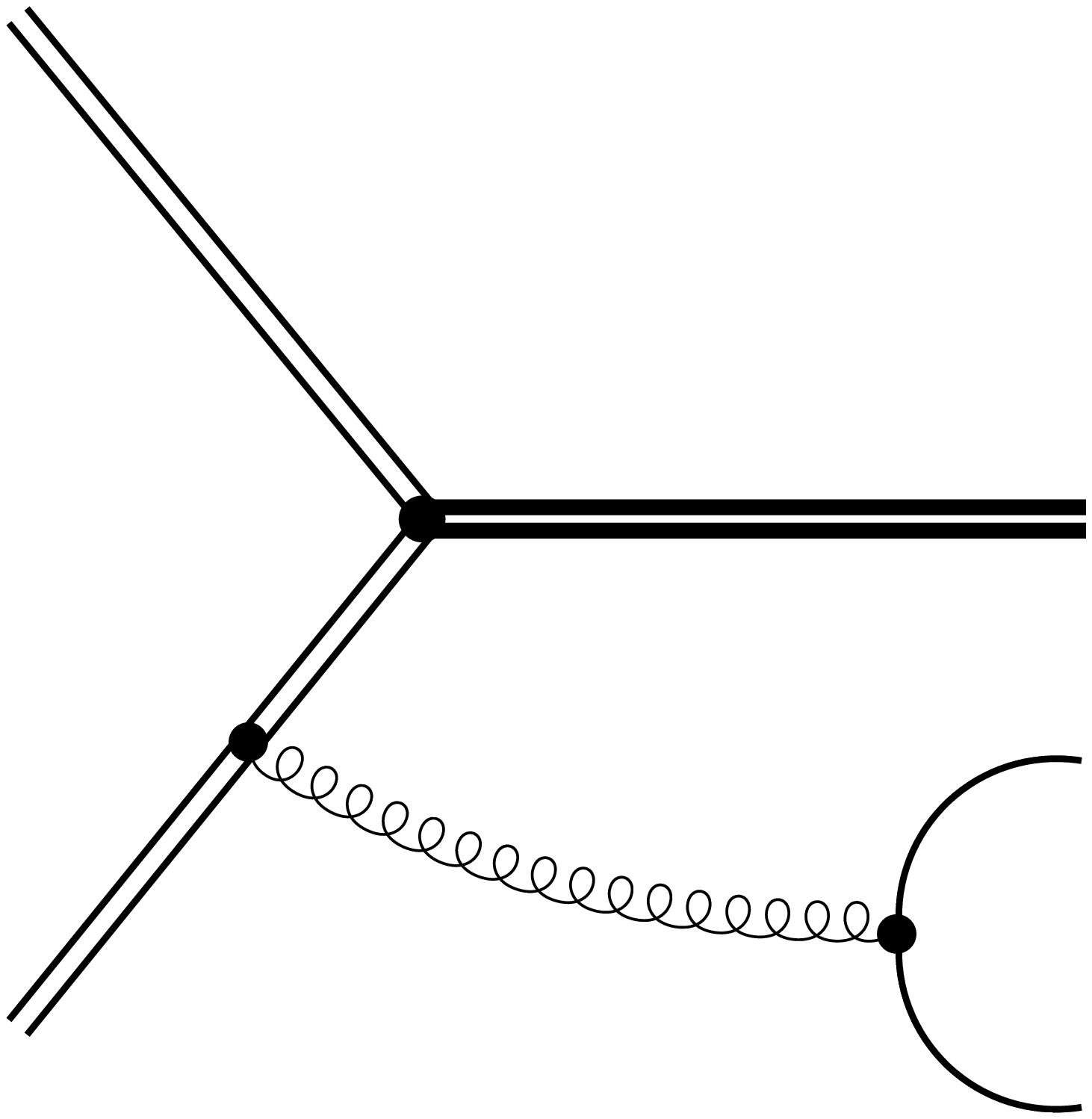}&\includegraphics[width=29mm,angle=0]{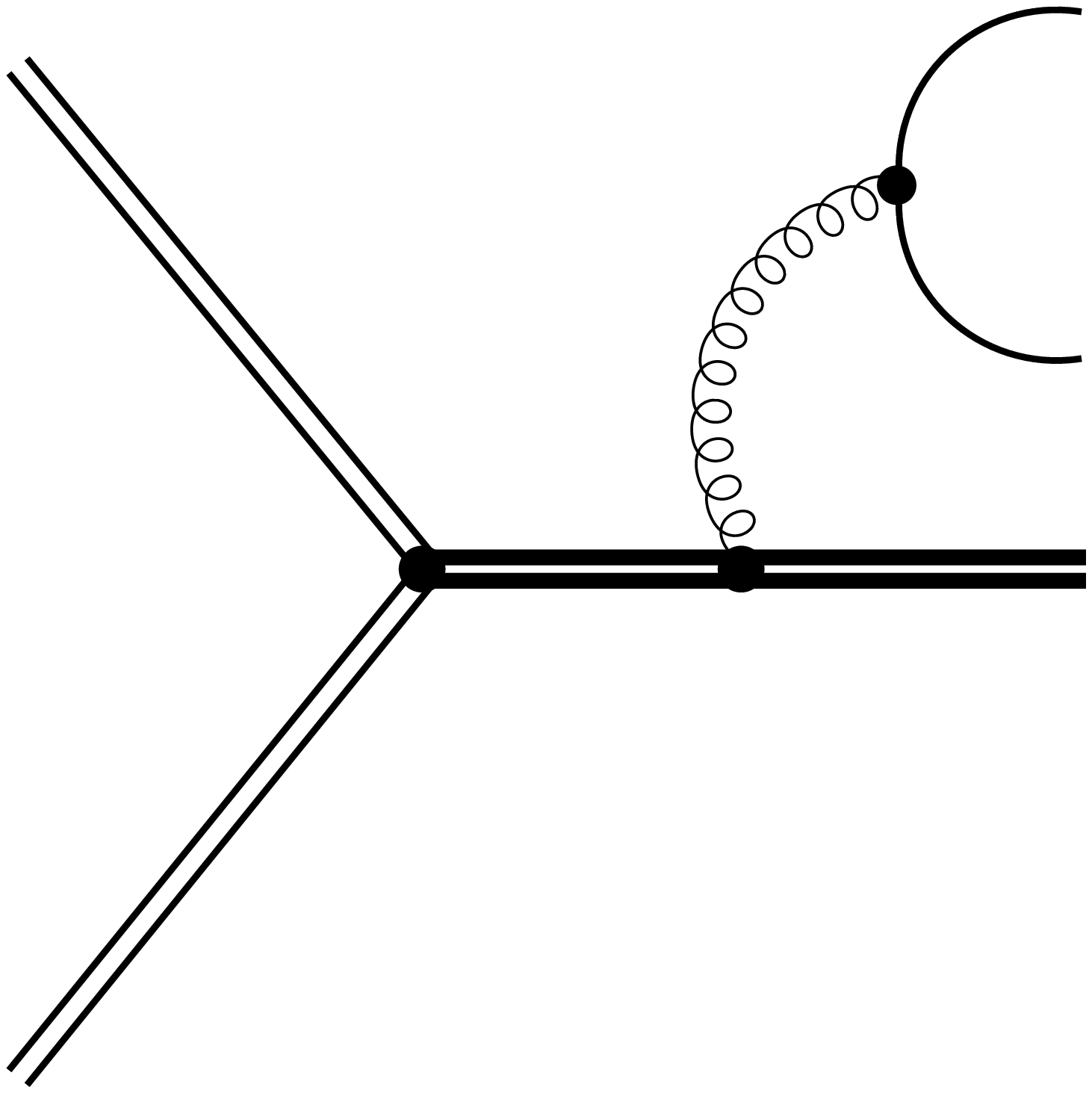}\\
\hline
\end{tabular}
\end{center}
\caption{\sf Complete set of double-real emission graphs contributing
  to the soft function. The graphs are divided into: (A)
  two emissions from eikonal lines, (B) gluon splitting
  after emission from an eikonal line, (C) massless quark-pair
  emission. \label{fig:EikonalRR}}
\end{figure}

\noindent
The complete set of double-real emission graphs is depicted in
Fig.~\ref{fig:EikonalRR}. We have divided the diagrams into three
categories: (A) two emissions from eikonal lines, (B)
gluon splitting after emission from an eikonal line, (C) massless
quark-pair emission. The division into (A) and (B) is not gauge
invariant. In order to uniquely define the contributions of the
interferences between the (A) and (B) categories, we mention that we
work in the Feynman gauge and that we take the gluon 
polarization sums to be
\begin{equation}
\sum_\lambda \epsilon_\mu(q,\lambda) \epsilon^*_\nu(q,\lambda)
\rightarrow -g_{\mu\nu} \; .
\end{equation}
In consequence, we also have to take ghost pairs into account, when
evaluating the square of the sum of the diagrams from (B).

In our calculation, we have not made use of non-abelian exponentiation
\cite{Mitov:2010rp, Gardi:2013ita} (see, however, next subsection for
an application in the real-virtual case). The color factors are
obtained for $SU(N_c)$ and subsequently translated into Casimir
operators.  The occurring phase space integrals can be evaluated
analytically by the following method. After canceling numerators with
denominators where possible, partial fractioning is used to obtain
denominators with the smallest number of different scalar products
from the set $n_i \cdot q_j$, $n_i \cdot (q_1 + q_2)$, $q_1 \cdot
q_2$, where $q_1$ and $q_2$ are the momenta of the emitted
gluons. Scalar products of the gluon momenta with $v$ are harmless,
since they only depend on the gluon energy. Subsequently, the
denominators involving $n_i \cdot (q_1 + q_2)$, are split with a
Mellin-Barnes representation
\begin{equation}
\frac{1}{(n_i \cdot (q_1+q_2))^\alpha} = \frac{1}{\Gamma(\alpha)}
\int_{-i \infty}^{+i \infty} \frac{dz}{2\pi i} \, \frac{1}{(n_i
  \cdot q_1)^{\alpha+z}(n_i \cdot q_2)^{-z}} \Gamma(\alpha+z) \Gamma(-z) \; ,
\end{equation}
where the contour is chosen to separate the poles of the gamma
functions. The angular integrations can be performed with the following
formulae \cite{Somogyi:2011ir}
\begin{equation}
\begin{split}
\int d\Omega_{{d-1}}(q)\,\frac{(q^0)^\alpha}{(n_{{i}} \cdot q)^\alpha}&=2^{2-\alpha-2\epsilon}\,
\pi^{1-\epsilon}\,\frac{\Gamma(1-\alpha-\epsilon)}{\Gamma(2-\alpha-2\epsilon)}
\; , \\ \\
\int d\Omega_{{d-1}}(q)\,\frac{(q^0)^{\alpha+\beta}}{(n \cdot q)^\alpha(\bar n \cdot q)^\beta}
&=2^{2-\alpha-\beta-2\epsilon}\,\pi^{1-\epsilon}\,
\frac{\Gamma(1-\alpha-\epsilon)\,\Gamma(1-\beta-\epsilon)}{\Gamma(1-\epsilon)
\Gamma(2-\alpha-\beta-2\epsilon)} \; , \\ \\
\int d\Omega_{{d-1}}(q_2)\,\frac{(q_1^0)^\beta(q_2^0)^{\alpha+\beta}}{(n_i \cdot q_2)^\alpha(q_1 \cdot q_2)^\beta}
&=2^{2-\alpha-\beta-2\epsilon}\,\pi^{1-\epsilon}
\,\frac{1}{\Gamma(\alpha)\,\Gamma(\beta)
  \,\Gamma(2-\alpha-\beta-2\epsilon)}\\
&\!\!\!\! \!\!\!\! \!\!\!\! \times\int_{-i\infty}^{+i\infty}\frac{dz}{2\pi
  i}\,\Gamma(-z)\Gamma(z+\alpha) \Gamma(z+\beta)
\Gamma(1-\alpha-\beta-\epsilon-z) \left(\frac{n_i \cdot q_1}{2 q_1^0}
\right)^z \; .
\end{split}
\end{equation}
The energy integrations in the resulting integrals can be
performed directly in terms of the Euler beta function
\begin{equation}
\int dE_1 \, dE_2 \, \delta(\omega-E_1-E_2) \, E_1^\alpha \, E_2^\beta
= \frac{\Gamma(\alpha+1)\Gamma(\beta+1)}{\Gamma(\alpha+\beta+2)}
\omega^{\alpha+\beta+1} \; .
\end{equation}
After application of the Barnes' lemmas, we end up with at most
one-fold Mellin-Barnes integrals, which can be resummed to
hypergeometric functions by closing contours and taking residues. In
practice, we have used the packages {\sc MB} \cite{Czakon:2005rk} and
{\sc HypExp} \cite{Huber:2005yg} for the manipulation of Mellin-Barnes
integrals and hypergeometric functions respectively.

\begin{figure}[h]
\begin{center}
\includegraphics[width=50mm,angle=0]{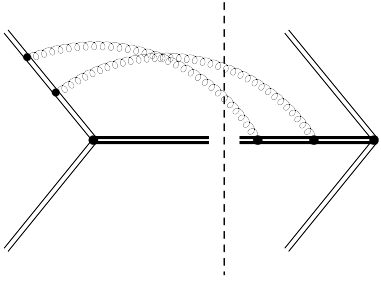}
\caption{\sf Interference diagrams contributing a hypergeometric
  function to $s_{{\Square}}^{(2)}$. \label{fig:Box}}
\end{center}
\end{figure}

The results are given separately for three parts.
The first part corresponds to the square of category (A) from
Fig.~\ref{fig:EikonalRR}. The result contains a single hypergeometric
function, due to the interference diagrams shown in
Fig.~\ref{fig:Box}. We obtain
\begin{equation}
\begin{split}
s_{{\Square}}^{(2)}&=-e^{2\gamma_{{E}}\epsilon}\frac{\Gamma^2(1-\epsilon)}{\Gamma(1-4\epsilon)}\\
                               &\times\left(C_A^2\,\left(\frac{2-19\epsilon+64\epsilon^2-74 \epsilon^3}{12 (1-2 \epsilon)^2\,(1-4 \epsilon)\,\epsilon^3 }+\frac{\Gamma (1+\epsilon)\,\Gamma (1+2 \epsilon)\,\Gamma (1-2 \epsilon)\,\Gamma (1-3 \epsilon)}{24 \epsilon^3\,\Gamma^2(1-\epsilon)}\right.\right.\\
                               &\left.\left.-\frac{1}{12 \epsilon^2
  (1-2 \epsilon)}\,_3 F_2(1,1-2 \epsilon,1-\epsilon;2-2
\epsilon,1+\epsilon;1)\right)-\,C_A\,C_R\,\frac{3-14 \epsilon}{4
  \epsilon^3\,(1-2 \epsilon)}+C_R^2\,\frac{2}{\epsilon^3}\right) \; .
\end{split}
\end{equation} 

The second part, due to the interference of  categories (A) and (B)
is the most complicated. The result contains several hypergeometric
functions 
\footnote{A subset of the graphs evaluated here occurs in the case of
  singlet production, i.e. for the soft function for the Drell--Yan process
  given in \cite{Belitsky:1998tc}. The result from
  \cite{Belitsky:1998tc} contains an Appell function, which can be
  expressed in terms of a ${}_3F_2$ hypergeometric function:
  $F_2(1,1+\epsilon,-2\epsilon,2+\epsilon,1-2\epsilon;1,1)\,=
  \frac{1+\epsilon}{2\epsilon}\left(\frac{2\,\Gamma(1-\epsilon)\,
    \Gamma^3(1+\epsilon)}{\Gamma(1+2\epsilon)}-\,_3
  F_2(1,1-\epsilon,-2 \epsilon;1-2 \epsilon,1-2 \epsilon;1)\right)$.}
\begin{equation}
\begin{split}
s_{{\bigtriangleup}}^{(2)}&=-e^{2\gamma_{{E}}\epsilon}\frac{\Gamma^2(1-\epsilon)}{\Gamma(1-4\epsilon)}\\ &\times\left(C_A^2\left(\frac{3-\epsilon+2
  \epsilon^2}{24\epsilon^3\,(1-2 \epsilon)^2
}+\frac{(1+\epsilon)\,\Gamma (1+\epsilon)\,\Gamma (1+2 \epsilon)
  \Gamma (1-2 \epsilon) \Gamma (1-3 \epsilon) }{12 \epsilon^3\,(1-2
  \epsilon)
  \Gamma^2(1-\epsilon)}\right.\right.\\ &\left.\left.+\frac{2-\epsilon}{24
  \epsilon^3\,(1-2 \epsilon)}\,_3 F_2(1,1-\epsilon,-2
\epsilon;1-2\epsilon,\epsilon;1)-\frac{1}{8 \epsilon^3}\,_3
F_2(1,1-\epsilon,-2 \epsilon;1-2 \epsilon,1-2
\epsilon;1)\right.\right.\\ &\left.\left.-\frac{1-\epsilon}{4
  \epsilon^2\,(1-2 \epsilon)^2}\,_3 F_2(1,1-2 \epsilon,2-\epsilon;2-2
\epsilon,1+\epsilon;1)\right)+\,C_A\,C_R\,\left(\frac{1-\epsilon}{2
  \epsilon^3\,(1-2 \epsilon)}\right.\right.\\ &\left.\left.+\frac{1}{4
  \epsilon^3}\,_3 F_2(1,1-\epsilon,-2 \epsilon;1-2 \epsilon,1-2
\epsilon;1)\right)\right) \; .
\end{split}
\end{equation} 

The third part is given by the sum of the squares of categories (B)
and (C). It is particularly simple, because it can be thought of as
the ${\cal O}(\alpha_s)$ contribution with an insertion of the
imaginary part of the gluon vacuum polarization on the gluon line. The
result reads
\begin{equation}
\begin{split}
s_{{\Circle}}^{(2)}&=-e^{2\gamma_{{E}}\epsilon}\frac{\Gamma^2(1-\epsilon)}{(3-2\epsilon)\,(1-2\epsilon)\,\Gamma(1-4\epsilon)}\\
                               &\times\left(C_A^2\,\frac{5-3 \epsilon}{4 \epsilon\,(1-4 \epsilon)}+\,C_A\,C_R\,\frac{5-3 \epsilon}{4 \epsilon^2}-\,C_A\,T_F\,n_f\,\frac{1-\epsilon}{\epsilon
 \,(1-4\epsilon)}-\,C_R\,T_F\,n_f\,\frac{1-\epsilon}{\epsilon^2}\right)
\; .
\end{split}
\end{equation} 

\subsection{Real-virtual corrections}

\noindent
Let us now consider virtual corrections to the soft function. Since we
are working in dimensional regularization, a Feynman integral does not
vanish if the result of the integration can be represented in the form
\begin{equation}
\int d^d k f(k) \propto \Lambda^{a+b \epsilon} \; ,
\end{equation}
where $a$ and $b$ are some real constants with $b \neq 0$, $\Lambda$
is a dimensionful parameter, and the proportionality coefficient does
not contain any other dimensionful parameters with $\epsilon$
dependent exponents. In the case of purely virtual corrections
involving external eikonal lines only, the result must have integer
scaling with respect to all the available momentum parameters. This
implies the vanishing of $b$, since the denominators of the Feynman
integrands contain no other dimensionful parameters. In consequence,
purely virtual corrections vanish.

Non-vanishing contributions are obtained once at least one real
particle, a gluon, is emitted, as for the real-virtual corrections we
wish to discuss in this subsection. In this case, we have two possible
types of dimensionful parameters, which are invariant under hard
momentum rescaling, and can thus be used in place of $\Lambda$ above
\begin{equation}
\Lambda_1(p,q) = \frac{m}{p \cdot q} \; , \;\; \Lambda_2(p_1,p_2,q) =
\frac{p_1 \cdot p_2}{p_1 \cdot q \; p_2 \cdot q} \; ,
\end{equation}
where $q$ is the momentum of the external gluon, while $p$ with $p^2 =
m^2$, $p_1$ and $p_2$ are the momenta of the eikonal lines.

\begin{figure}[h]
\begin{center}
\begin{tabular}{|cc|cc|}
\multicolumn{1}{c}{\includegraphics[width=29mm,angle=0]{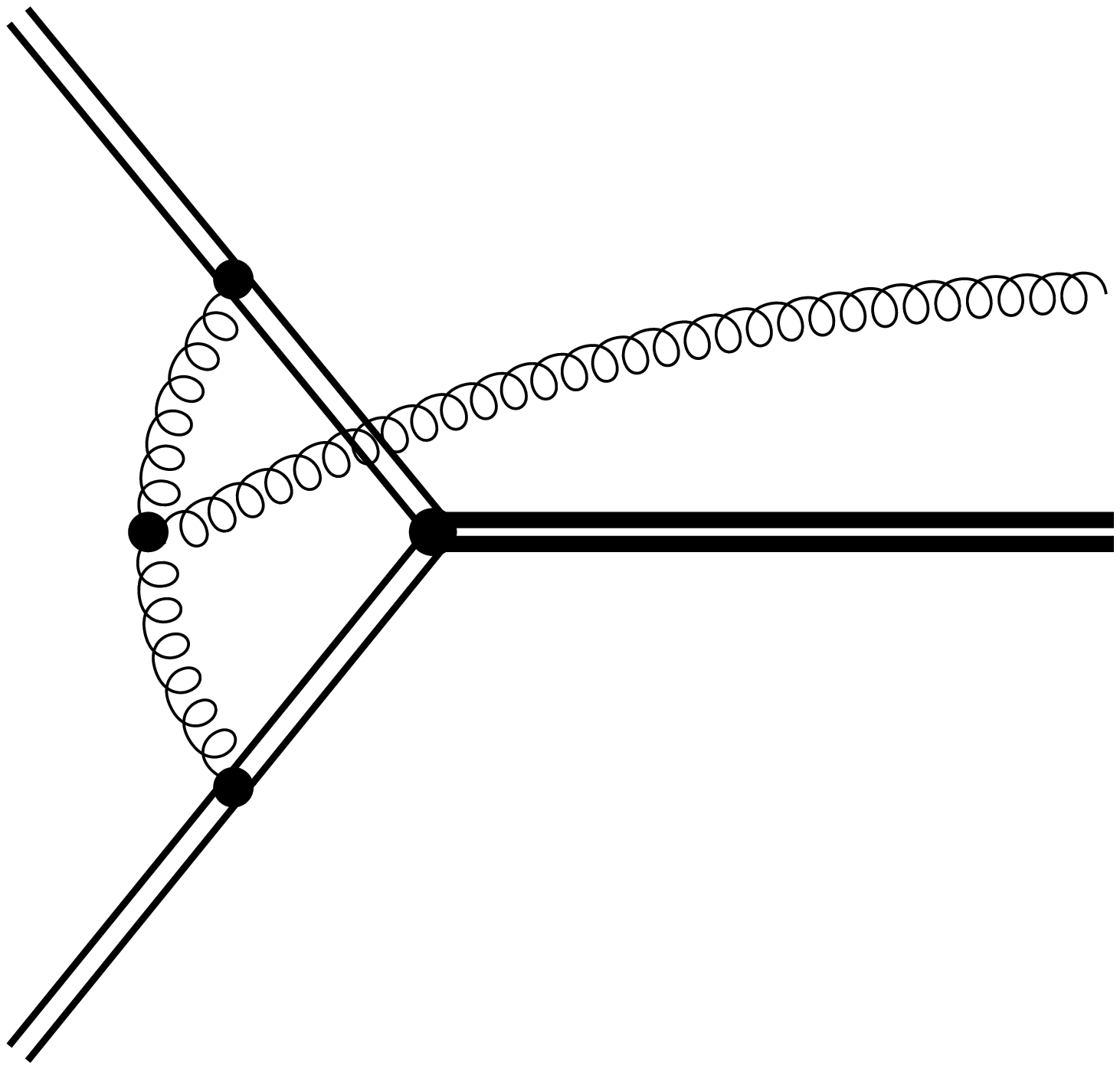}}&\multicolumn{1}{c}{\includegraphics[width=29mm,angle=0]{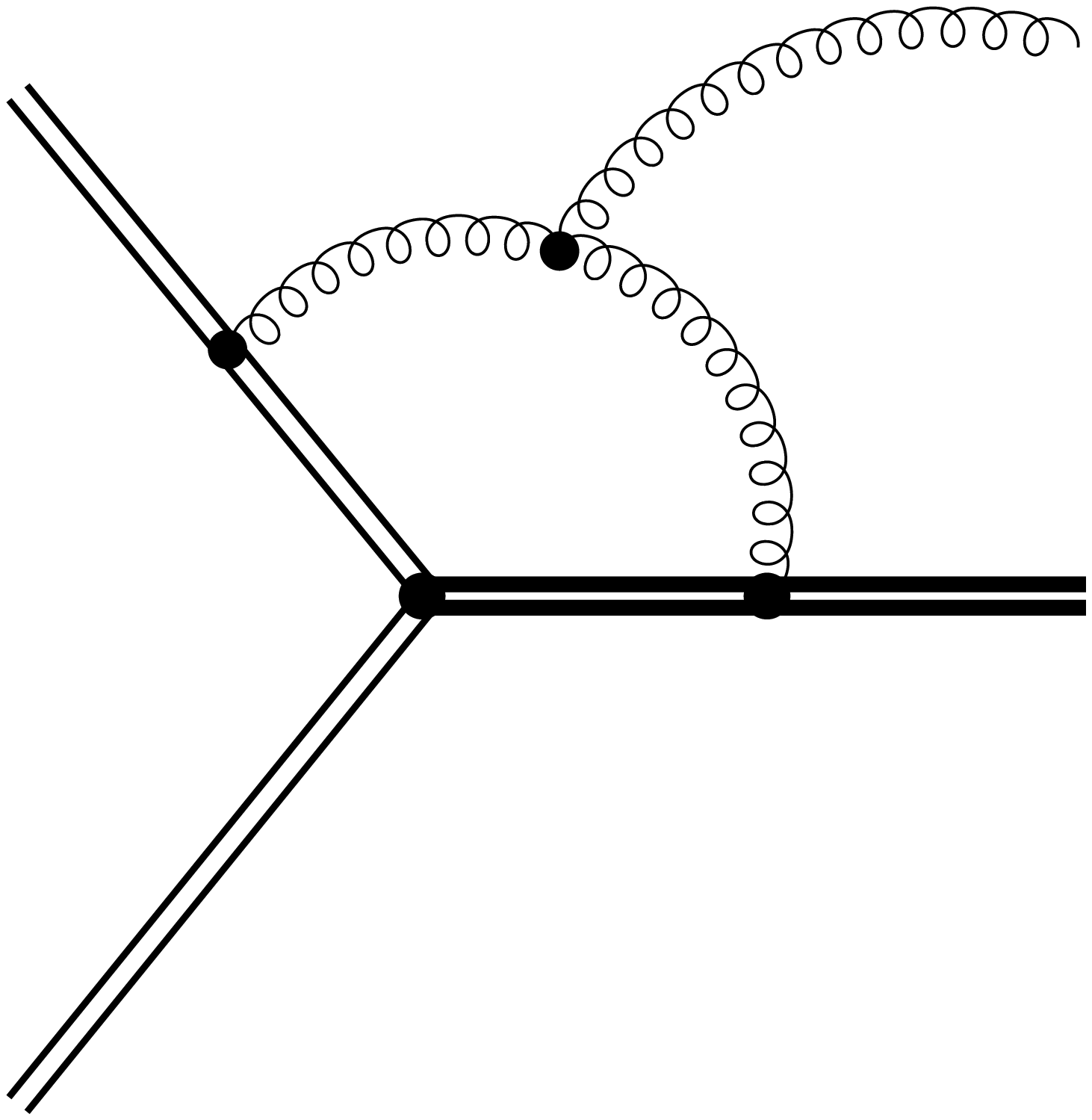}}&\multicolumn{1}{c}{\includegraphics[width=29mm,angle=0]{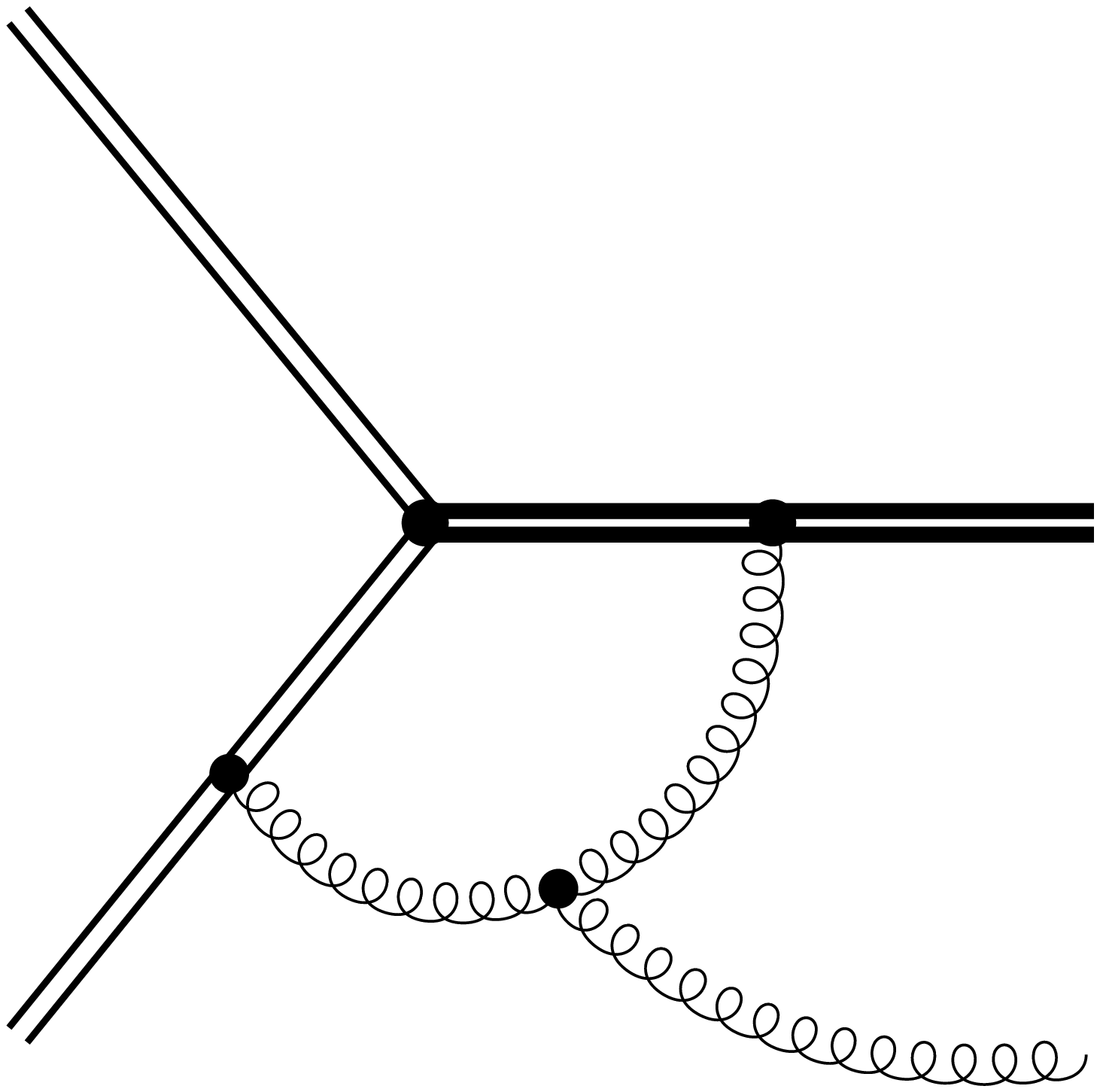}}&\multicolumn{1}{c}{\includegraphics[width=29mm,angle=0]{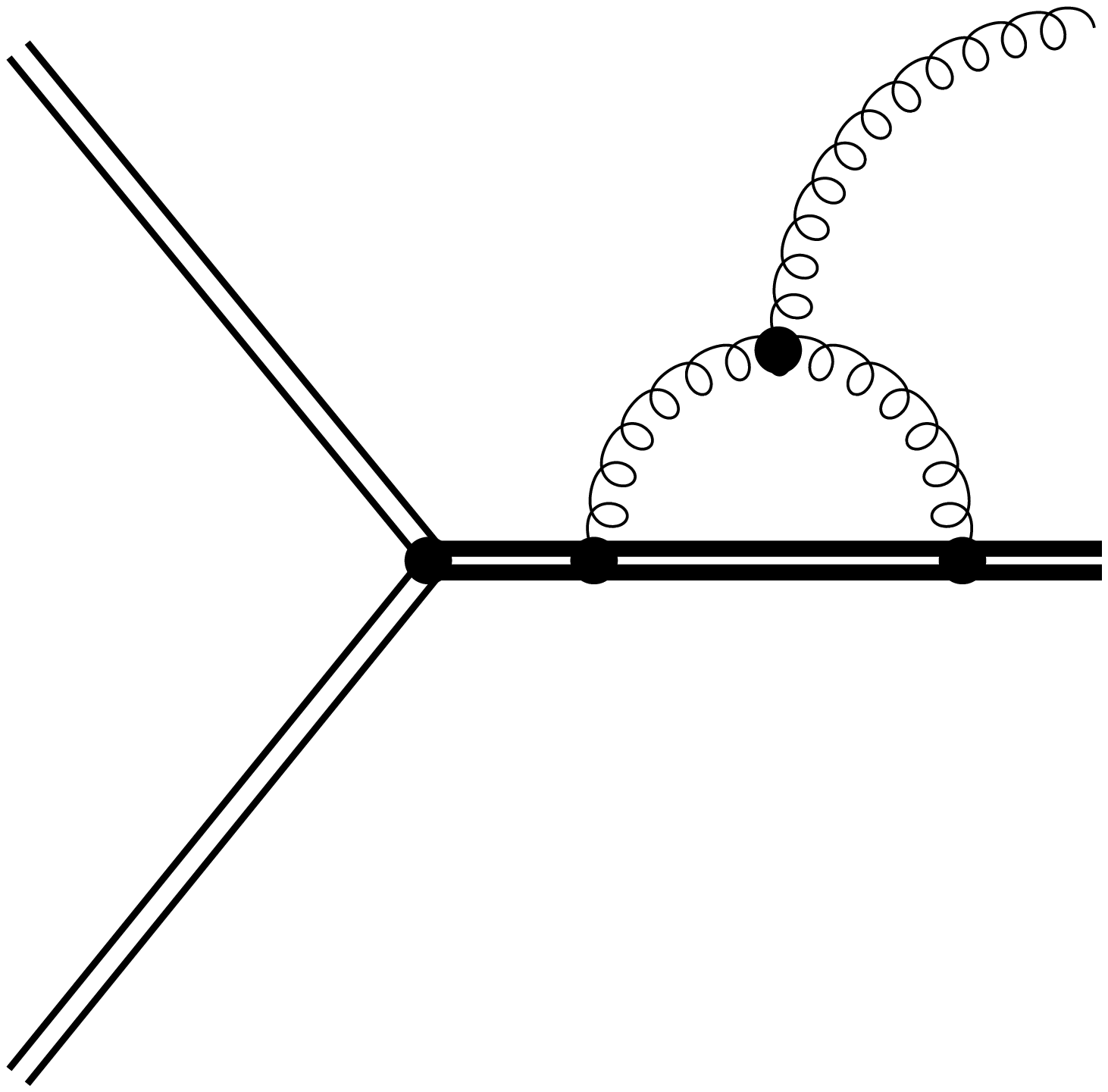}}\\
\hline
&&&\\
\includegraphics[width=29mm,angle=0]{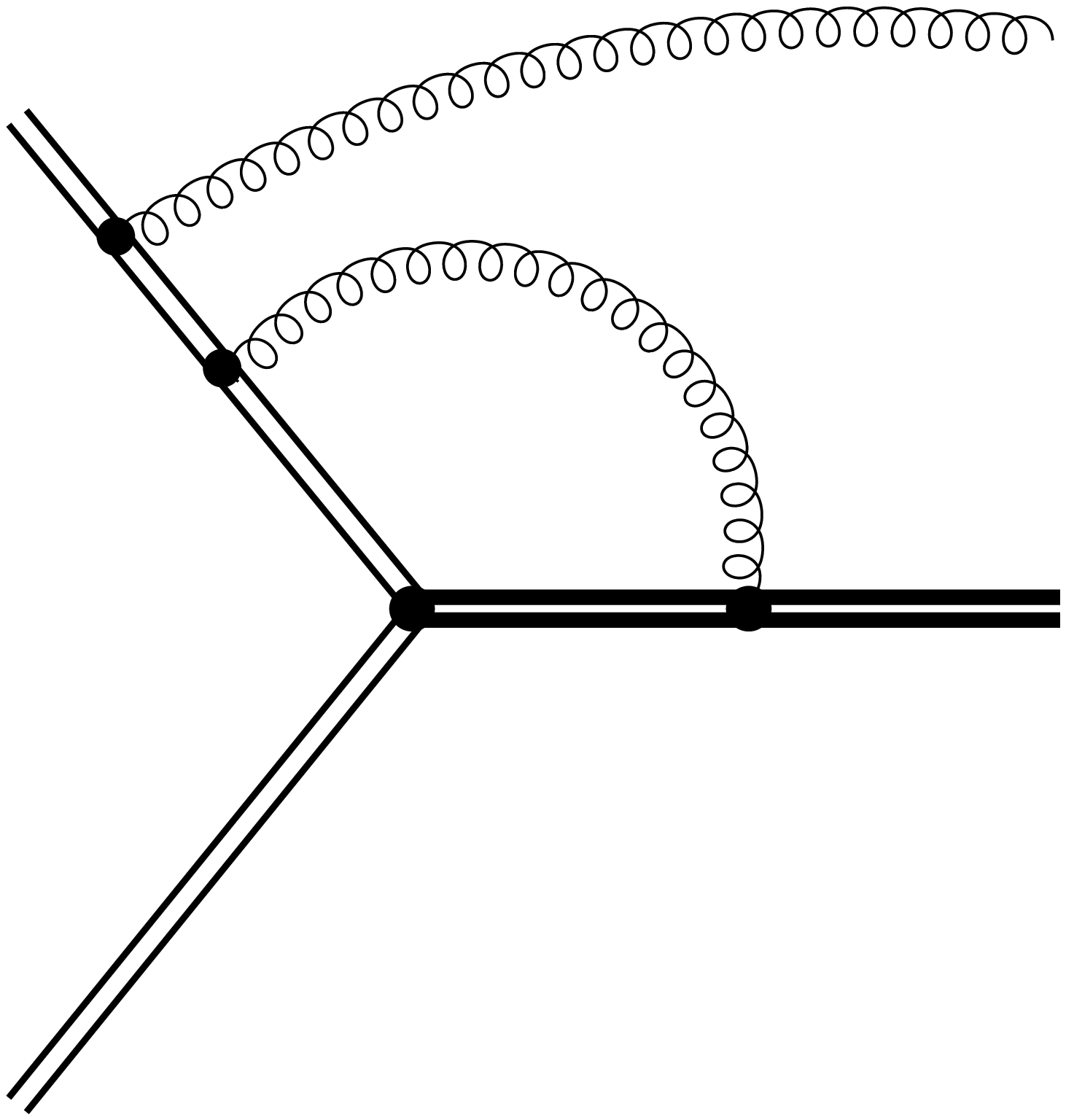} &\includegraphics[width=29mm,angle=0]{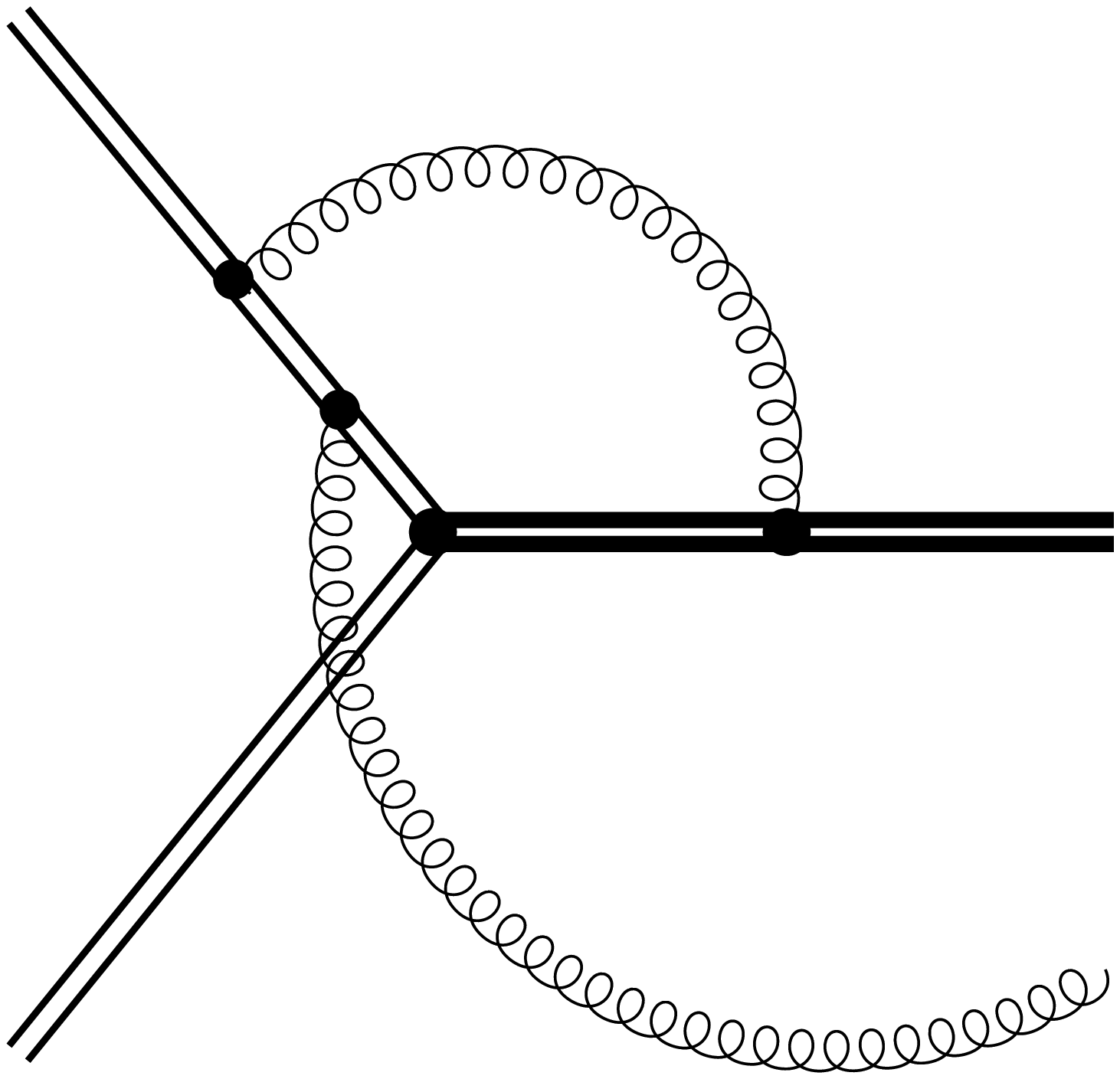}&\includegraphics[width=29mm,angle=0]{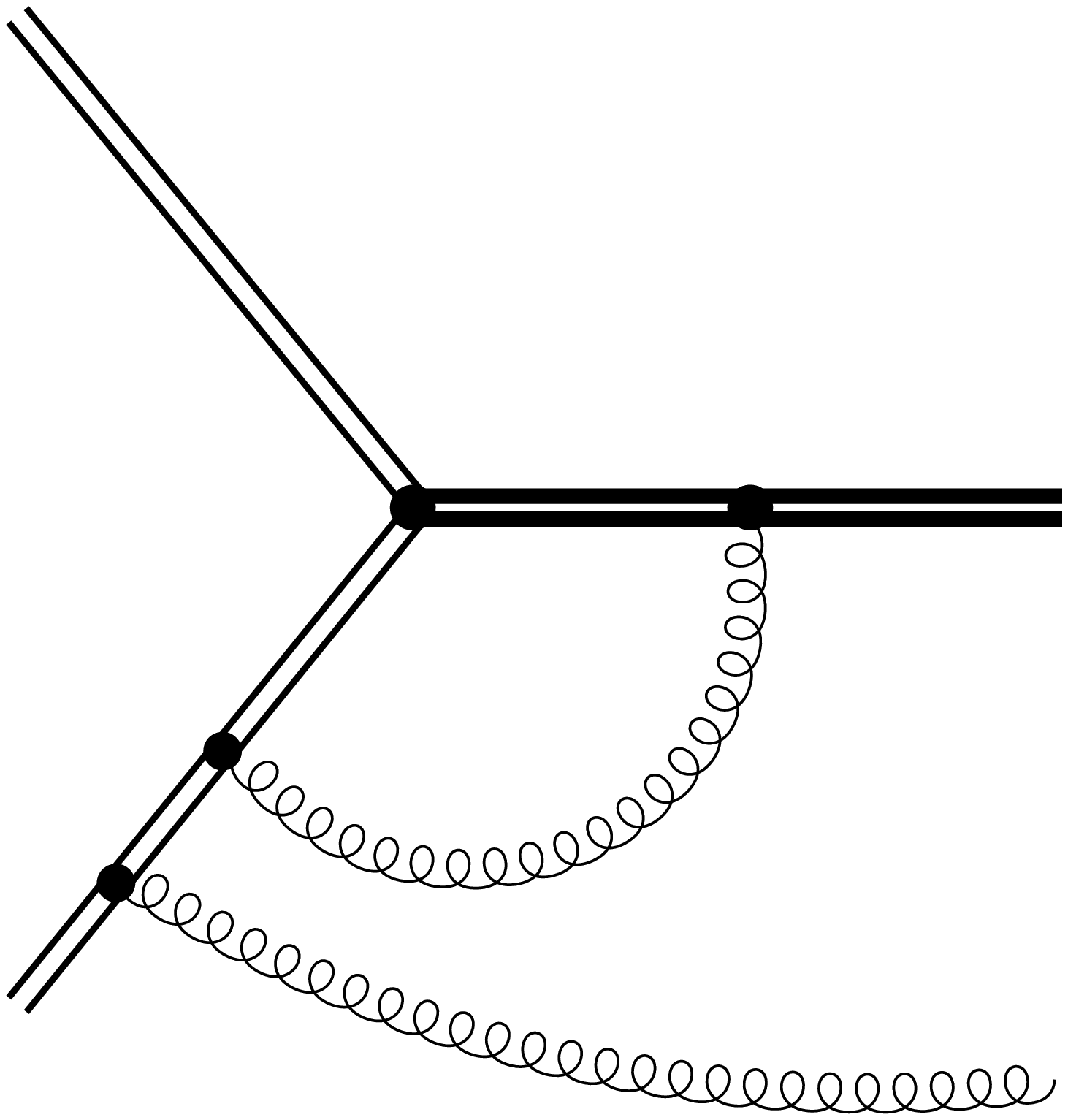}&\includegraphics[width=29mm,angle=0]{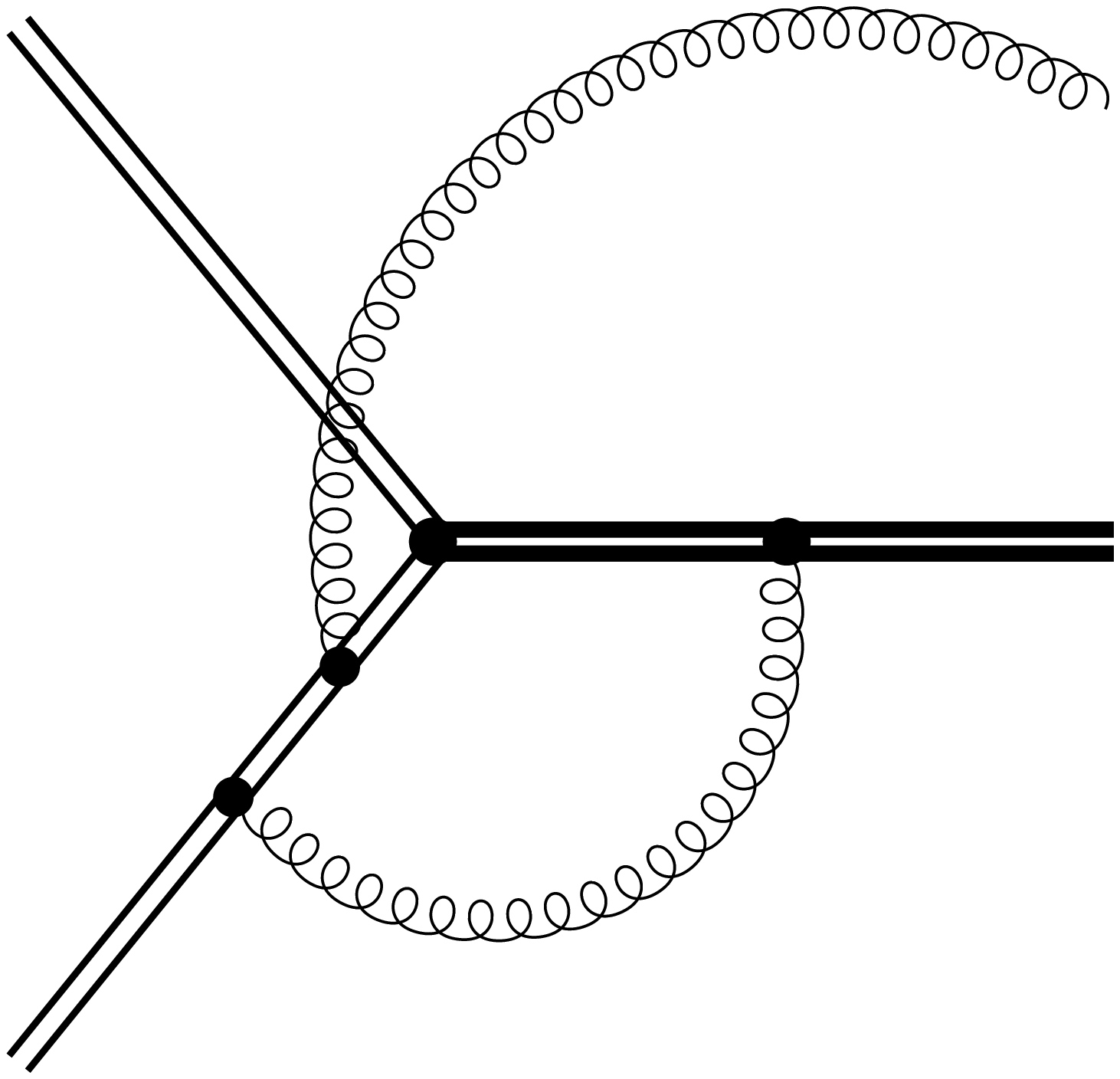}\\
\hline
&&&\\
\includegraphics[width=29mm,angle=0]{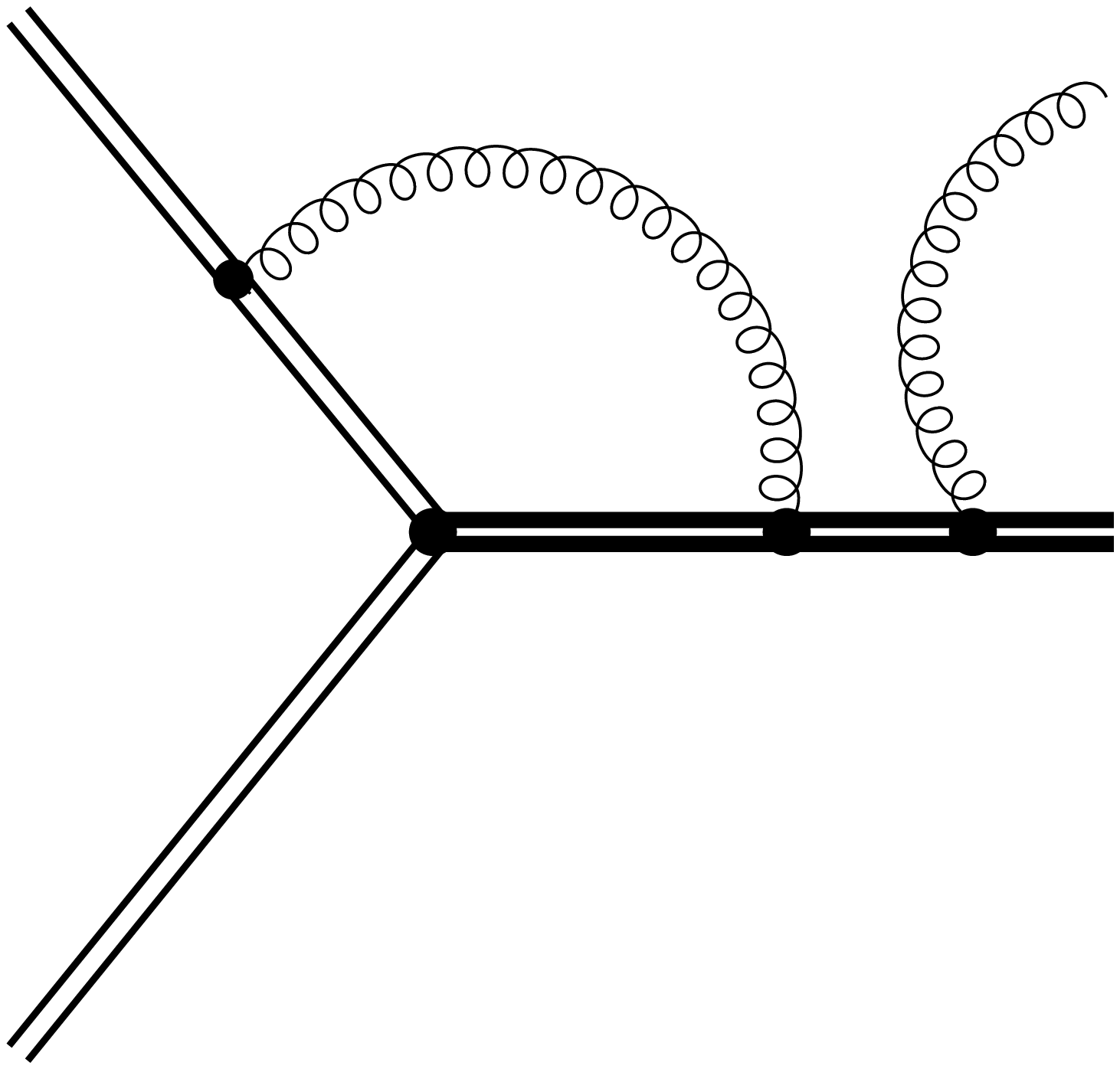} &\includegraphics[width=29mm,angle=0]{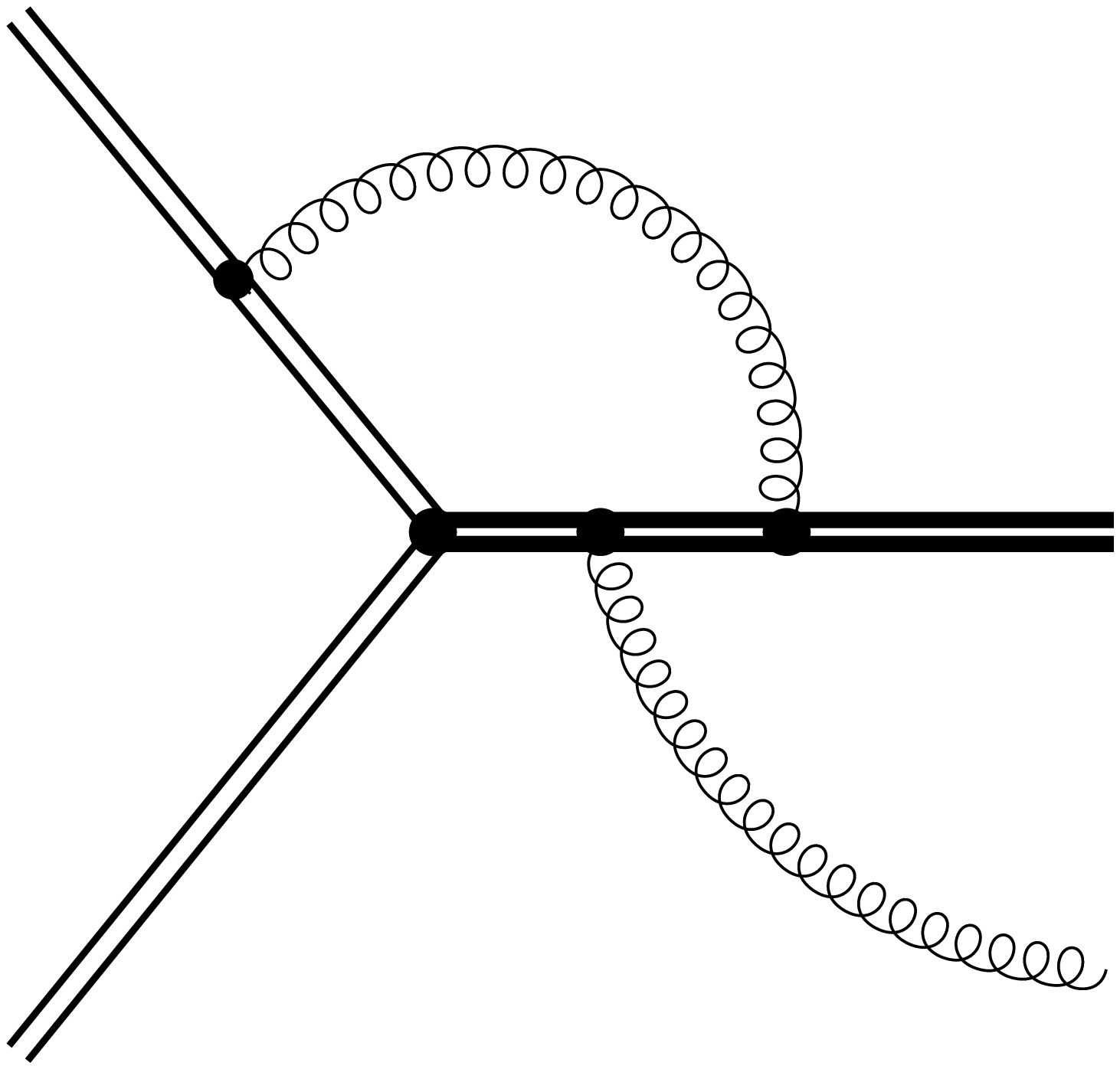}&\includegraphics[width=29mm,angle=0]{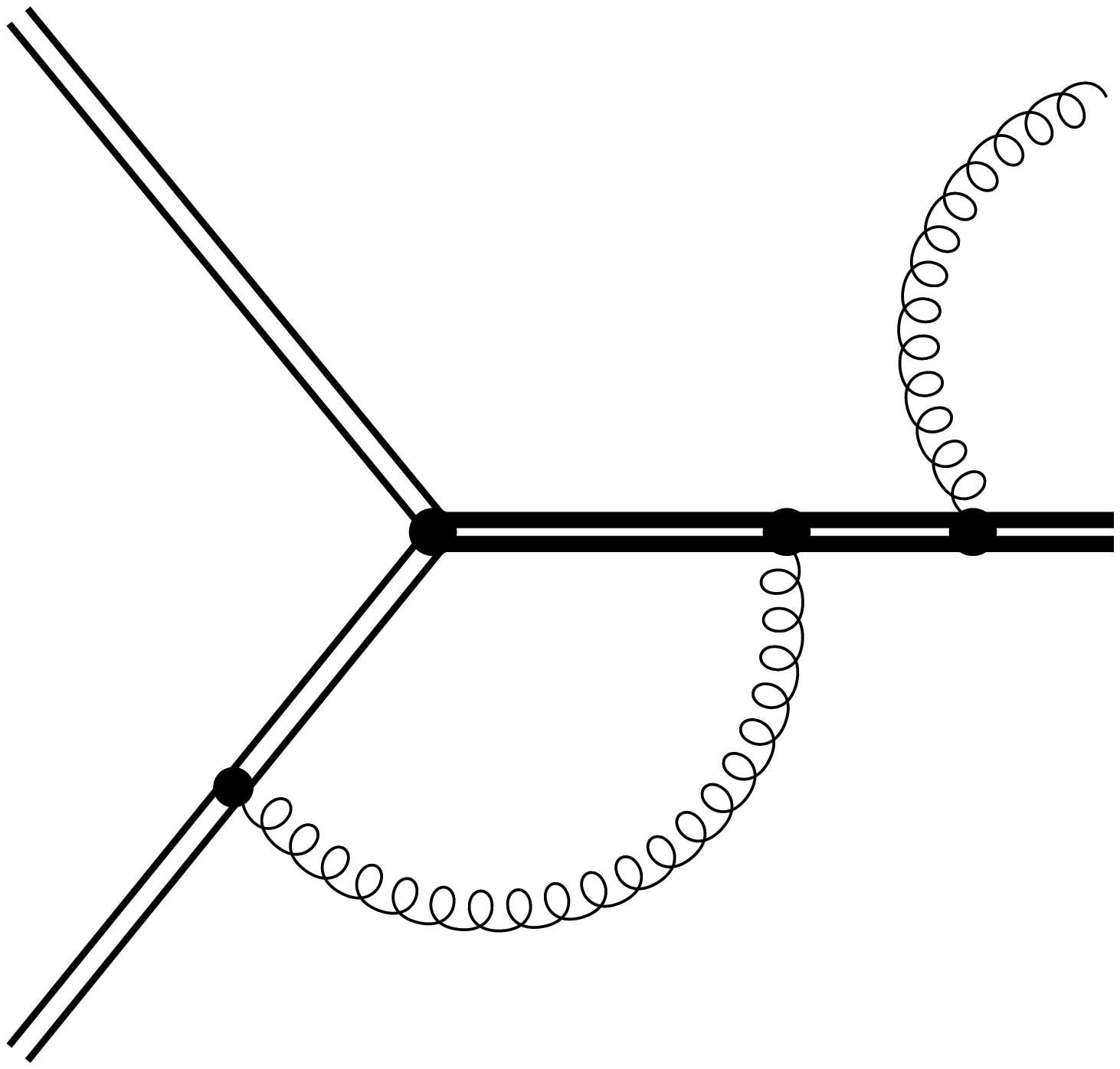}&\includegraphics[width=29mm,angle=0]{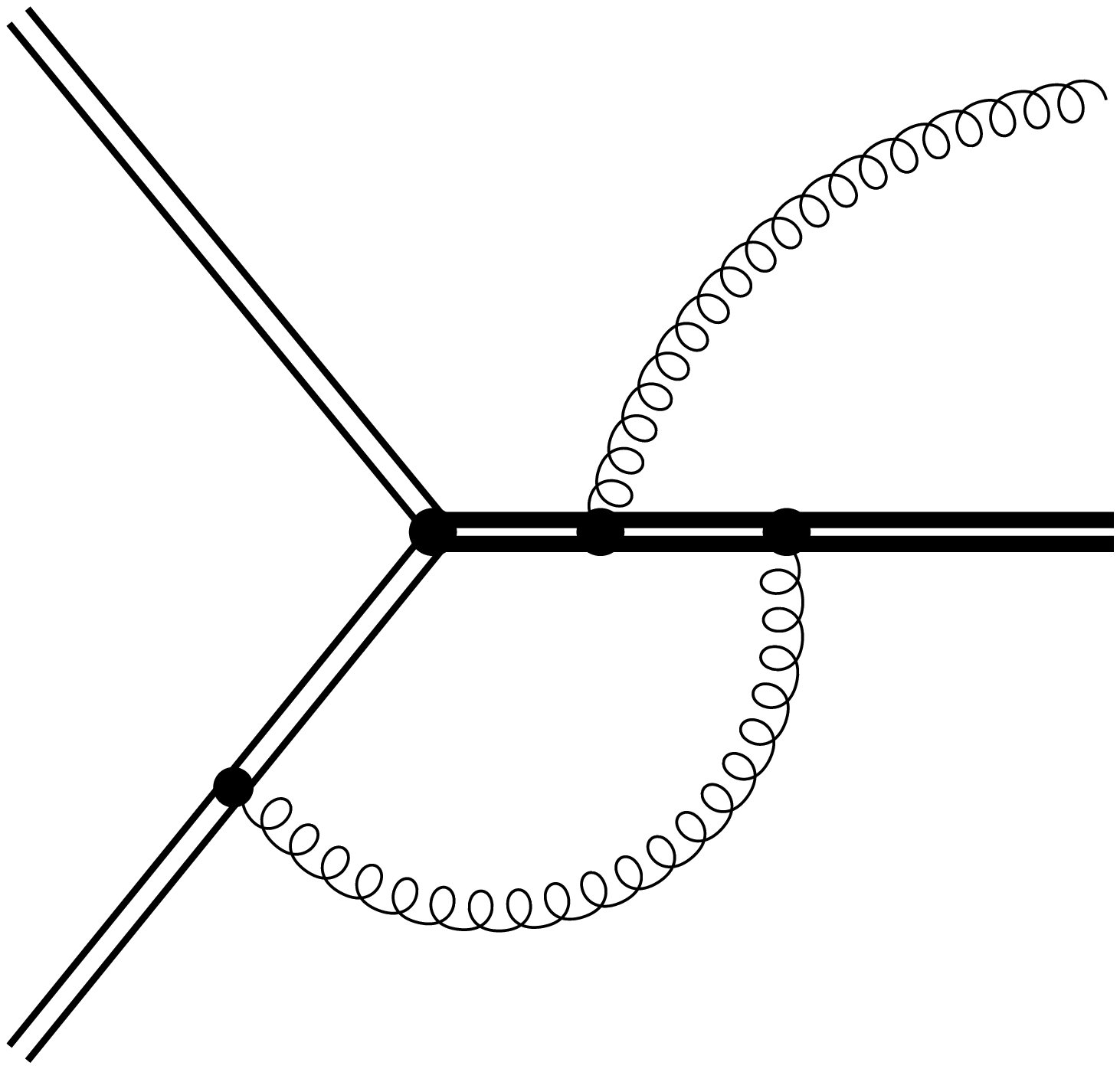}\\
\hline
&&\multicolumn{2}{c}{}\\
\includegraphics[width=29mm,angle=0]{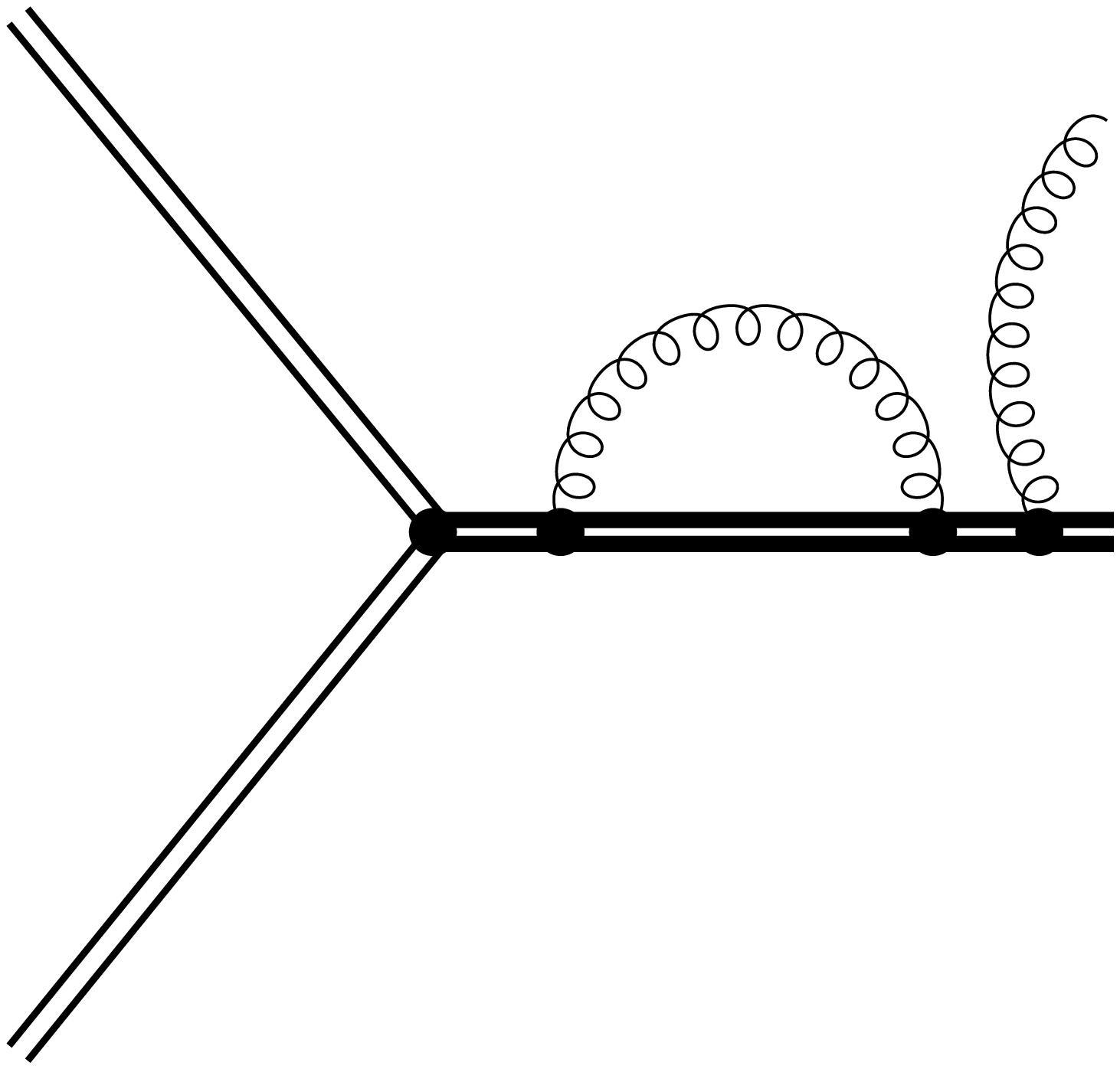} &\includegraphics[width=29mm,angle=0]{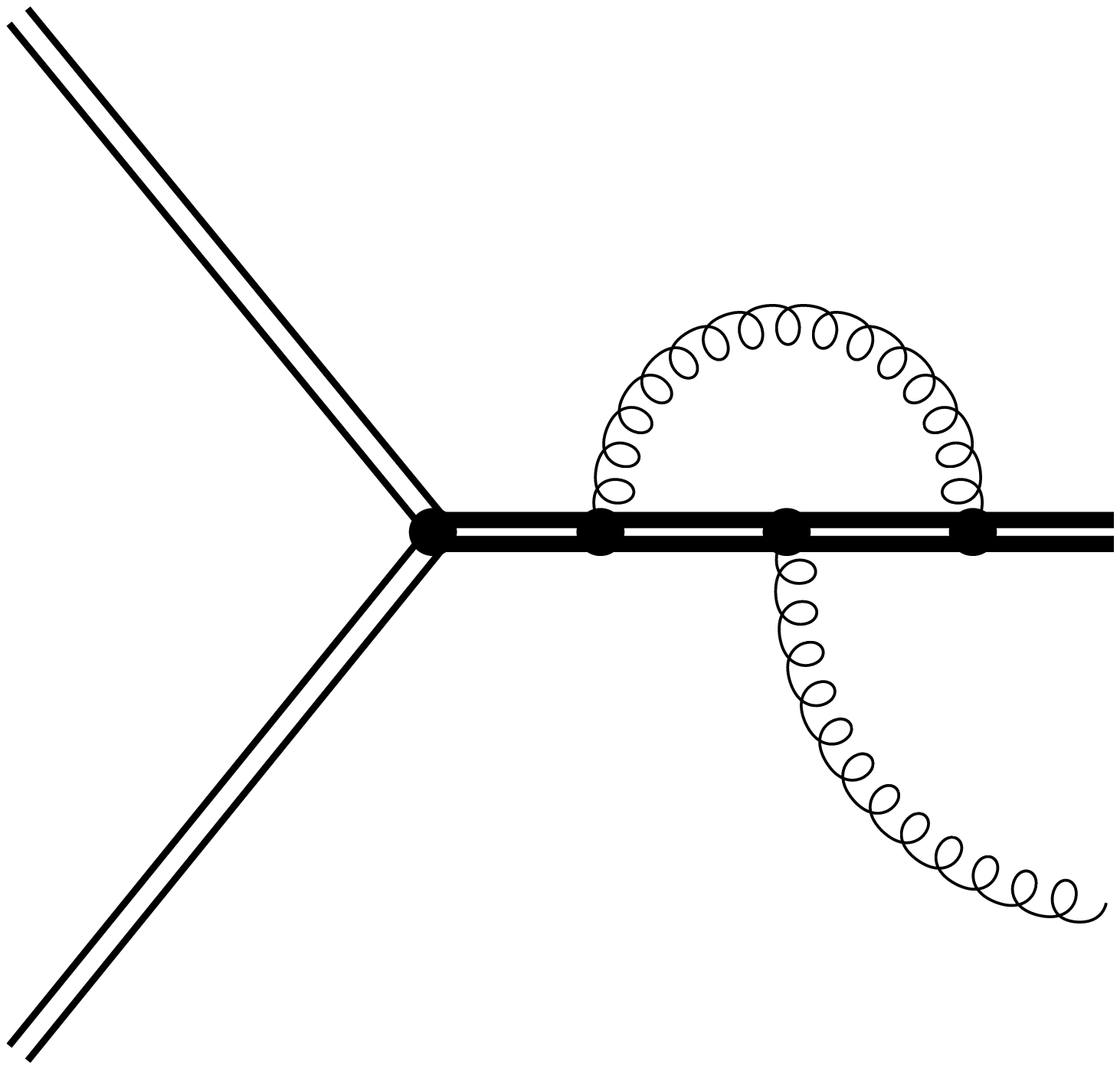}&\multicolumn{2}{c}{}\\
\cline{1-2}
\end{tabular}
\end{center}
\caption{\sf Complete set of non-vanishing real-virtual graphs
  contributing to the soft function. The pairs of graphs in boxes can
  be combined as explained in the text. \label{tab:EikonalRV}}
\end{figure}

We can now consider two different types of emissions: from an eikonal
line, or from a virtual gluon line. In the case of an emission from an
eikonal line, the denominators of the Feynman integrands depend on the
emitted gluon momentum through $p_1 \cdot (k + q)$ only, with $p_1$,
$q$ and $k$ the momenta of the emitting eikonal line, the emitted
gluon, and the loop integration respectively. Therefore, the result of the
integration depends non-trivially only on $p_1 \cdot q$ (ignoring the
numerator, which does not influence the $d$-dimensional scaling of
dimensionful parameters), and thus either on $\Lambda_1(p_1,q)$ or
$\Lambda_2(p_1,p_2,q)/\Lambda_1(p_2,q)$, where $p_2$ is the momentum
of the other eikonal line integrated in the loop (if the integration
involves two different eikonal lines). Such emissions will
thus only give non-vanishing  contributions, if either $p_1^2
\neq 0$ or $p_2^2 \neq 0$. Emissions from gluon
lines give non-vanishing contributions either if the loop contains two
different eikonal lines, in which case the result can be expressed through
$\Lambda_2(p_1,p_2,q)$, or if the single eikonal line in the loop is
massive, in which case the result is expressed through
$\Lambda_1(p,q)$. The possible graphs generated according to these
considerations are shown in Fig.~\ref{tab:EikonalRV}.

\begin{figure}
\begin{center}
\includegraphics[width=80mm,angle=0]{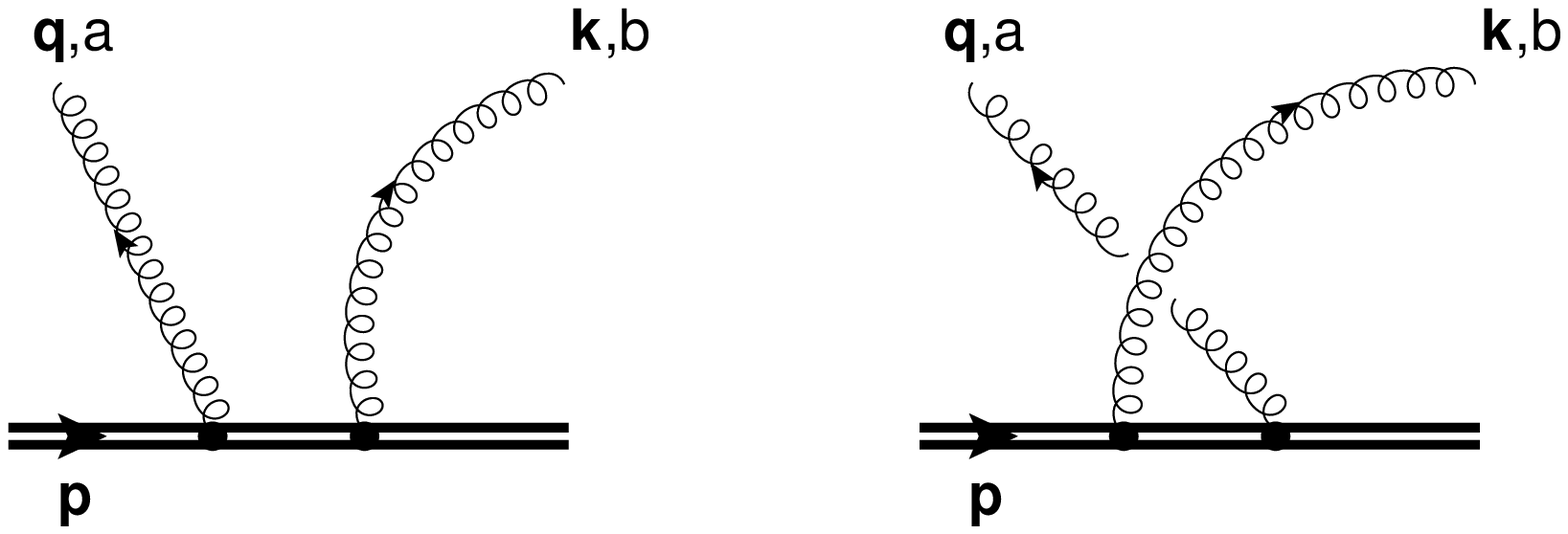}
\end{center}
\caption{\sf Real gluon emission (momentum $q$ and color index $a$) from
  an eikonal line in the presence of a virtual gluon (momentum $k$ and
  color index $b$).\label{fig:EikonalId}}
\end{figure}

The pairs of graphs in boxes in Fig.~\ref{tab:EikonalRV} can be
combined using color algebra and eikonal identities. This is a simple
case of eikonal exponentiation \cite{Mitov:2010rp, Gardi:2013ita} for
multiple eikonal lines joined at one point. Consider for
example the two graphs of Fig.~\ref{fig:EikonalId}. The color factor
of the right graph can be represented as the sum of the color factor
of the left graph and an additional contribution
\begin{equation}\label{colorid}
{\bf T}^{(\bm R) \, a} \, {\bf T}^{(\bm R) \, b} =
{\bf T}^{(\bm R) \, b} \, {\bf T}^{(\bm R) \, a} + if^{abc} {\bf T}^{(\bm
  R) \, c} \; .
\end{equation}
If we ignore the second term on the right hand side of this equation,
the two graphs will only differ in the kinematics. Their sum can then
be written under the same integral sign and will contain the factor
\begin{equation}
\frac{1}{p \cdot q} \frac{1}{p \cdot (k + q)} + 
\frac{1}{p \cdot k} \frac{1}{p \cdot (k + q)} =
\frac{1}{p \cdot q} \frac{1}{p \cdot k} \; .
\end{equation}
This is the basic eikonal identity, which shows that the two emissions
can be  completely factorized. Such a contribution corresponds to a
product of a real emission and a pure virtual contribution. It thus
vanishes by the arguments from the beginning of this subsection. In
consequence, for each pair of emissions as in
Fig.~\ref{fig:EikonalId}, it is sufficient to only consider the second
graph with a modified color factor corresponding to the second term on
the right hand side of Eq.~(\ref{colorid}).

As far as the virtual integrals are concerned, it turns out that they
are the same as those that were calculated in
\cite{Bierenbaum:2011gg}. The complete list is
\begin{center}
\begin{tabular}{llll}
$M_1$ : &$\int d^{d}k\,\frac{1}{[k^2]\,[(k+q)^2]\,[n_i\cdot k]}$ \;
  ,&$\int d^{d}k\,\frac{1}{[k^2]\,[(k+q)^2]\,[-v\cdot k]}$ \, ,&\\
&&&\\
$M_2$ : &$\int d^{d}k\,\frac{1}{[k^2]\,[-n_i\cdot k-n_i\cdot
      q]\,[-v\cdot k]}$ \, , & $\int d^{d}k\,\frac{1}{[k^2]\,[v\cdot
      k+v\cdot q]\,[n\cdot k]}$ \, , & \!\!\!\! \!\!\!\! \!\!\!\! $\int
  d^{d}k\,\frac{1}{[k^2]\,[v\cdot k+v\cdot q]\,[-v\cdot k]}$ \, , \\
&&&\\
$M_3$ : &$\int d^{d}k\,\frac{1}{[k^2]\,[(k+q)^2]\,[-n\cdot k-n\cdot
      q]\,[\overline{n}\cdot k]}$ \, , & $\int
  d^{d}k\,\frac{1}{[k^2]\,[(k+q)^2]\,[v\cdot k-v\cdot q]\,[n_i\cdot
      k]}$ \, , &\\ 
\end{tabular}
\end{center}
where the naming of the master integrals, $M_1$, $M_2$ and $M_3$,
matches that of Ref.~\cite{Bierenbaum:2011gg}. Our integrals are all integrated to
Euler gamma functions in $d$-dimensions, because we only have one massive
line.

The remaining phase space integration has the same complexity as in the
${\cal O}(\alpha_s)$ case. We use a subset of the formulae from the
calculation of the double-real corrections, but never have to
introduce any Mellin-Barnes integrations. Therefore, the final result
does not contain any hypergeometric functions and is given by
\begin{equation}
\begin{split}
s_{{\CIRCLE}}^{(2)}&=\frac{e^{2\gamma_{{E}}\epsilon}}{\Gamma(1-4\epsilon)}\left(C_{{A}}^2\,\left(\frac{\Gamma (1+ \epsilon) \Gamma (1+2 \epsilon) \Gamma (1-2 \epsilon) \Gamma (1-3\epsilon)}{8
   \epsilon^3\,(1-2 \epsilon)}-\frac{\Gamma^3(1+\epsilon) \Gamma^3(1-\epsilon)}{4 \epsilon^3\,\Gamma (1+2 \epsilon)}\right.\right.\\
                               &\left.\left.+\frac{\left(1-2\epsilon+4
  \epsilon^2\right)\,\Gamma^2(1+2\epsilon) \Gamma^2(1- \epsilon)}{8
  \epsilon^3 (1-2 \epsilon)^2 \Gamma (1+4
  \epsilon)}\right)+C_{{A}}\,C_{{R}}\frac{\Gamma^3(1+\epsilon)
  \Gamma^3(1-\epsilon)}{2 \epsilon^3\,\Gamma (1+2 \epsilon)}\right) \; .
\end{split}
\end{equation} 

\subsection{Renormalized result}

\noindent
After summing the contributions from the previous subsections and
renormalizing according to Eq.~(\ref{eq:SR}), we obtain the
following perturbative expansion of the soft function for color octet
production at threshold in Mellin space
\begin{equation}
S(L) = 1 + \frac{\alpha_s}{\pi} \, S^{(1)}(L)
 + \left(\frac{\alpha_s}{\pi}\right)^2 \, S^{(2)}(L) + {\cal
   O}(\alpha_s^3) \; ,
\end{equation}
with
\begin{equation}
\begin{split}
S^{(1)}(L)&= C_{{A}} (L+1)+C_{{R}} \left(2 L^2+\frac{\pi
  ^2}{12}\right) \; , \\
S^{(2)}(L) &= C_{{A}}^2 \left(\frac{17 L^2}{12}+\left(\frac{\zeta (3)}{2}+\frac{151}{36}-\frac{\pi ^2}{12}\right)L-\frac{5 \zeta (3)}{8}+\frac{13 \pi ^4}{2880}+\frac{\pi
   ^2}{24}+\frac{223}{54}\right)\\
              &+C_{{R}}\;C_{{A}} \left(\frac{29 L^3}{9}+\left(\frac{103}{18}-\frac{\pi ^2}{6}\right) L^2+ \left(-\frac{7 \zeta (3)}{2}+\frac{101}{27}+\frac{\pi
   ^2}{12}\right)L\right.\\
              &\left.-\frac{11 \zeta (3)}{72}-\frac{\pi ^4}{48}+\frac{139 \pi ^2}{864}+\frac{607}{324}\right)\\
              &+C_{{R}}^2 \left(2 L^4+\frac{\pi ^2 L^2}{6}+\frac{\pi ^4}{288}\right)\\
              &+C_{{A}}\;T_{{F}}\;n_f \left(-\frac{L^2}{3}-\frac{11 L}{9}-\frac{40}{27}\right)\\
              &+C_{{R}}\;T_{{F}}\;n_f \left(-\frac{4 L^3}{9}-\frac{10 L^2}{9}-\frac{28 L}{27}+\frac{\zeta
   (3)}{18}-\frac{5 \pi ^2}{216}-\frac{41}{81}\right) \; . 
\end{split}
\end{equation}
In case the initial state is a color octet as well, i.e. $C_R = C_A$,
this result further simplifies to become
\begin{equation}
\begin{split}
S_{{A}}^{(1)}(L)&= C_{{A}} \left(2 L^2+L+\frac{\pi ^2}{12}+1\right) \;
, \\
S_{{A}}^{(2)}(L) &= C_{{A}}^2 \left(2 L^4+\frac{29 L^3}{9}+\frac{257 L^2}{36}+\left(\frac{857}{108}-3 \zeta (3)\right)L-\frac{7 \zeta (3)}{9}-\frac{37 \pi ^4}{2880}+\frac{175 \pi
   ^2}{864}+\frac{1945}{324}\right)\\
              &+C_{{A}}\;T_{{F}}\;n_f \left(-\frac{4 L^3}{9}-\frac{13
  L^2}{9}-\frac{61 L}{27}+\frac{\zeta (3)}{18}-\frac{5 \pi
  ^2}{216}-\frac{161}{81}\right) \; .
\end{split}
\end{equation}
Of course, the logarithmic terms proportional to $L^n$, $n>0$, in
these expressions can be derived directly from the RGE
Eq.~(\ref{eq:RGE}). The new results are thus at $L=0$.

%%%%%%%%%%%%%%%%%%%%%%%%%%%%%%%%%%%%%%%%%%%%%%%%%%%%%%%%%%%%%%%%%%%%%%%%%%%%%%%%

\section{Conclusions and outlook}

\noindent 
We have presented the result for the next-to-next-to-leading order
soft function for color octet production at threshold. The primary use
of this result is the derivation of the constants in the threshold
expansion of NNLO QCD cross section for heavy-flavor
pair-production. This will be one of the topics covered in an upcoming
publication.

Our results may have further applications, for example in the
determination of threshold expansions of NNLO cross sections
for other massive particle (e.g. squarks, gluinos) production in pairs,
or as single fundamental states (e.g. color octet scalars). Essential is only
the combined color configuration of the final state. The results could
also be generalized to arbitrary representations, since the most
complicated part of the calculation was the evaluation of the
integrals. We have refrained from such a generalization in this
publication.

Since the result for the bare soft function is exact in
$d$-dimensions, it can be used as part of a calculation of the soft
function at higher orders.

%%%%%%%%%%%%%%%%%%%%%%%%%%%%%%%%%%%%%%%%%%%%%%%%%%%%%%%%%%%%%%%%%%%%%%%%%%%%%%%%

\section*{Acknowledgments}

\noindent 
This research was supported by the German Research
Foundation (DFG) via the Sonderforschungsbereich/Transregio SFB/TR-9
``Computational Particle Physics''. The work of M.C. was supported by
the DFG Heisenberg programme.

%%%%%%%%%%%%%%%%%%%%%%%%%%%%%%%%%%%%%%%%%%%%%%%%%%%%%%%%%%%%%%%%%%%%%%%%%%%%%%%%

\appendix

%%%%%%%%%%%%%%%%%%%%%%%%%%%%%%%%%%%%%%%%%%%%%%%%%%%%%%%%%%%%%%%%%%%%%%%%%%%%%%%%

\section{Wilson lines and their properties}
\label{sec:wilson}

\noindent
We consider the following Wilson line operator
\begin{equation}\label{wilson}
\bm{\Phi}^{(\bm R)}_\beta(x;b,a) = P \exp\left( i g_s^0 \int_a^b
dt\, \beta \cdot A^c(x+t \,\beta)\, {\bf T}^{(\bm R) \, c} \right) \; ,
\end{equation}
which represents the contribution to the path integral of a classical
particle charged under the representation $\bm{R}$ of the gauge group,
and moving in a straight line with four-momentum $\beta$ between the
points $x + a\beta$ and $x + b\beta$. The operator acts in color
space, with $P$ in front of the exponential denoting path
ordering. This allows to account for the color state evolution of the
particle due to gluon emission. Finally, $g_s^0$ stands for the bare
coupling constant. The relation between the bare and renormalized
couplings is given at the level of $\alpha^0_s = (g^0_s)^2/4\pi$ as
\begin{equation}
\alpha_s^{0}\,=\left(\frac{e^{\gamma_E}}{4\pi}\right)^\epsilon\,\mu^{2\epsilon}\,
Z_{\alpha_s} \,\alpha_s \; .
\end{equation} 
The $\overline{\mbox{MS}}$ renormalization constant $Z_{\alpha_s}$ can
be determined using the scale independence of $\alpha_s^0$ and the
renormalization group equation satisfied by $\alpha_s$
\begin{equation}
\frac{d\alpha_s}{d\ln\mu^2}\,=\,\beta \; ,
\end{equation} 
with $\beta = -(\alpha_s^2/4\pi) \, b_0+{\cal O}(\alpha_s^3)$, $b_0 =
11/3\,C_A-4/3\,T_F\, n_f$.

In our exposition we do not need the shift, $x$, of
the trajectory, and often explicitly write the color
indices. Therefore, we introduce the following simplified notation
\begin{equation}
\bm{\Phi}^{(\bm R)}_\beta(b,a) = \bm{\Phi}^{(\bm R)}_\beta(x = 0;b,a)
\; , \;\; \Phi^{(\bm R)}_{\beta,cd}(b,a) =  \left( \bm{\Phi}^{(\bm
  R)}_\beta(b,a) \right)_{cd} \; .
\end{equation}

The gauge group representations relevant to this study are: {\bf 3}
for quarks, $\bm{\bar{3}}$ for anti-quarks, and {\bf 8} for
gluons. For these three cases, there is
\begin{equation}\label{color}
({\bf T}^{({\bm 3}) \, a})_{bc} = T^a_{bc} \; , \;\;
({\bf T}^{({\bm{\bar{3}}}) \, a})_{bc} = -T^a_{cb} \; , \;\;
({\bf T}^{({\bm 8}) \, a})_{bc} = if^{bac} \; .
\end{equation}
Notice that
\begin{equation}
\left( \bm{\Phi}^{(\bm{3}) \, \dagger}_\beta(b,a) \right)_{cd} =
\bm{\Phi}^{(\bm{\bar{3}})}_{\beta,dc}(b,a) \; ,
\end{equation}
because the hermitian conjugation implies the same change of the order
of the color operators in the path ordering, as the different
definition of the operators in Eq.~(\ref{color}), while the field
operators commute. Gauge transformation properties of Green functions
involving Wilson lines can be verified using covariance
\begin{equation}\label{eq:covariance}
\bm{\Phi}^{(\bm{R})}_\beta(b,a)~\rightarrow~{\bf U}^{(\bm{R})}(b \beta)
\bm{\Phi}^{(\bm{R})}_\beta(b,a) {\bf U}^{(\bm{R}) \, \dagger}(a
\beta) \; ,
\end{equation}
where ${\bf U}^{(\bm{R})}(x)$ is the gauge transformation matrix at
point $x$ in the representation $\bm{R}$.

The Wilson line operator satisfies a first order linear differential
equation
\begin{equation}\label{diff0}
\frac{d}{dt} \bm{\Phi}^{(\bm R)}_\beta(t,a) = i g_s^0 \;
\beta \cdot A^c(t \,\beta)\, {\bf T}^{(\bm R) \, c} \;
\bm{\Phi}^{(\bm R)}_\beta(t,a) \; , \;\;
\bm{\Phi}^{(\bm R)}_\beta(a,a) = 1 \; ,
\end{equation}
where the second equation is the boundary condition. This leads to the
composition law
\begin{equation}
\bm{\Phi}^{(\bm R)}_\beta(c,b) \bm{\Phi}^{(\bm R)}_\beta(b,a) =
\bm{\Phi}^{(\bm R)}_\beta(c,a) \; ,
\end{equation}
which implies a differential equation on the second argument of the
operator
\begin{equation}\label{diff}
\frac{d}{dt} \bm{\Phi}^{(\bm R)}_\beta(a,t) = - \bm{\Phi}^{(\bm
  R)}_\beta(a,t) \; i g_s^0 \; \beta \cdot A^c(t \,\beta)\, {\bf T}^{(\bm
  R) \, c} \; ,
\end{equation}
which we shall use below. Since the operator acting on the right hand
side of the differential equation, Eq.~(\ref{diff0}), is anti-hermitian,
$\bm{\Phi}^{(\bm R)}_\beta(b,a)$ is unitary.

We restrict ourselves to Wilson lines for a straight line motion,
because they model soft gluon emission from a hard emitter, the
kinematics of which is not affected by the radiation. In momentum
space, soft radiation is described by eikonal lines, where each
emission contributes a kinematical factor  $p^\mu/p \cdot k$, with
$p$ and $k$ the hard and soft momenta respectively. This factor is
rescaling invariant in the hard momentum, a property which the Wilson
line must, therefore, also have. This can only be realized if the
arguments of $\bm{\Phi}^{(\bm R)}_\beta(b,a)$ are restricted to null or
infinity, i.e.\ $a,b~\in~\{-\infty,0,+\infty\}$. Our lines are
semi-infinite, i.e.\ start or end at 0, where the hard process takes
place in the semi-classical picture of the collision. In this case,
the integration in the exponent in the definition Eq.~(\ref{wilson})
is regulated by a factor $\exp(\pm \delta t)$, with $\delta
\rightarrow 0^+$, and sign as appropriate to make the integral
convergent. These factors correctly reproduce the causal $+i 0^+$
prescription in the eikonal propagators in momentum space.

Let us now consider the case of an out-going heavy-quark--anti-quark
pair at threshold. The quarks share the same four-momentum $v^\mu =
(1,\bm{0})$, where we have used rescaling invariance to rescale the
energy to unity. It has been shown in \cite{Beneke:2009rj} that as
long as the pair is in an irreducible representation of the gauge
group, the eikonal lines can be combined into one. We reproduce here
the argument in our notation for the singlet and octet cases.

For a singlet initial configuration after the production in the hard
process, the soft radiation is described by

\begin{equation}
\delta_{a'b'} \Phi^{(\bm{3})}_{v,aa'}(+\infty,0)
\Phi^{(\bm{\bar{3}})}_{v,bb'}(+\infty,0) = 
\left( \bm{\Phi}^{(\bm 3)}_v(+\infty,0) \bm{\Phi}^{(\bm
  3) \, \dagger}_v(+\infty,0) \right)_{ab} = \delta_{ab} \; ,
\end{equation}
which in fact means that there is no soft radiation from the final state. In
the case of an octet configuration we have to consider the combination
\begin{equation}
T^c_{a'b'} \Phi^{(\bm{3})}_{v,aa'}(+\infty,0)
\Phi^{(\bm{\bar{3}})}_{v,bb'}(+\infty,0) = 
\left( \bm{\Phi}^{(\bm 3)}_v(+\infty,0) \, {\bf T}^{({\bm 3}) \, c} \,
\bm{\Phi}^{(\bm 3) \, \dagger}_v(+\infty,0) \right)_{ab} \; .
\end{equation}
We wish to prove that this combination can be replaced by a single
line in the octet representation. This is equivalent to the statement
\begin{equation}
\bm{\Phi}^{(\bm 3)}_v(+\infty,t) \, {\bf T}^{({\bm 3}) \, a} \,
\bm{\Phi}^{(\bm 3) \, \dagger}_v(+\infty,t) = {\bf T}^{({\bm 3}) \, b} \,
  \Phi^{(\bm 8)}_{v,ba}(+\infty,t) \; ,
\end{equation}
at $t = 0$. The equation is trivially satisfied at $t = +\infty$. It is
also satisfied at all $t$, because both sides fulfill the same
differential equation obtained by taking a derivative in $t$ and using
Eq.~(\ref{diff}).

%%%%%%%%%%%%%%%%%%%%%%%%%%%%%%%%%%%%%%%%%%%%%%%%%%%%%%%%%%%%%%%%%%%%%%%%%%%%%%%%

\section{Anomalous dimensions}
\label{sec:AnomalousDimensions}

Here we give the anomalous dimensions necessary to determine the
renormalization constants for the hard functions. In the octet case,
they have been originally determined in 
Refs.~\cite{Czakon:2009zw,Beneke:2009rj} 
\begin{equation}
\begin{split}
\Gamma_H^{\bm{3}\otimes\bar{\bm 3}|\bm{1}}\left(\frac{\mu}{Q}\right) 
&=\frac{a_s}{\pi}\,C_F
\left(-4\,\ln\left(\frac{\mu}{Q}\right)-3\right)+
\left(\frac{a_s}{\pi}\right)^2\left(
C_F^2 \left(-6 \zeta(3)-\frac{3}{8}+\frac{\pi^2}{2}\right)\right.\\ 
&\left. +C_A\,C_F \left(\ln 
\left(\frac{\mu}{Q}\right)\left(\frac{\pi^2}{3} -\frac{67}{9}\right)
+\frac{13 \zeta(3)}{2}-\frac{11 \pi^2}{24}-\frac{961}{216}\right)\right.\\ 
&\left.+C_F\,T_F\,n_f
\left(\frac{20}{9} \ln \left(\frac{\mu }{Q}\right)+\frac{\pi^2}{6}+
\frac{65}{54}\right)\right) \; , \\
\Gamma_H^{\bm{3}\otimes\bar{\bm 3}|\bm{8}}\left(\frac{\mu}{Q}\right) 
&=\frac{a_s}{\pi}\left(C_F
\left(-4\,\ln\left(\frac{\mu}{Q}\right)-3\right)-C_A\right)+
\left(\frac{a_s}{\pi}\right)^2
\left(C_A^2 \left(-\frac{\zeta (3)}{2}-\frac{49}{36}+
\frac{\pi^2}{12}\right)\right.\\ &\left.+C_A\,C_F \left(\ln 
\left(\frac{\mu}{Q}\right)\left(\frac{\pi^2}{3} -\frac{67}{9}\right)+
\frac{13 \zeta(3)}{2}-\frac{11 \pi^2}{24}-\frac{961}{216}\right)\right.\\ 
&\left.+C_F^2 \left(-6 \zeta(3)-\frac{3}{8}+\frac{\pi^2}{2}\right)+
\frac{5}{9}\,C_A\,T_F\,n_f+C_F\,T_F\,n_f
\left(\frac{20}{9} \ln \left(\frac{\mu }{Q}\right)+\frac{\pi^2}{6}+
\frac{65}{54}\right)\right) \; , \\
\Gamma_H^{\bm{8}\otimes\bm{8}|\bm{1}}\left(\frac{\mu}{Q}\right)
&=\frac{a_s}{\pi} \left(C_A \left(-4\,\ln \left(\frac{\mu}{Q}\right)-
\frac{11}{3}\right)+\frac{4}{3}\,T_F\,n_f\right)\\ 
&+\left(\frac{a_s}{\pi}\right)^2
\left(C_A^2 \left(\ln \left(\frac{\mu }{Q}\right)\left(\frac{\pi^2}{3}
-\frac{67}{9}\right)+\frac{11 \pi^2}{72}+\frac{\zeta(3)}{2}-\frac{173}{27}\right)\right.\\ 
&\left.+C_A\,T_F\,n_f\left(\frac{20}{9} \ln \left(\frac{\mu }{Q}\right)-
\frac{\pi^2}{18}+\frac{64}{27}\right)+C_F\,T_F\,n_f\right) \; , \\
\Gamma_H^{\bm{8}\otimes\bm{8}|\bm{8_{A,S}}}\left(\frac{\mu}{Q}\right)
&=\frac{a_s}{\pi} \left(C_A \left(-4\,\ln \left(\frac{\mu}{Q}\right)-
\frac{14}{3}\right)+\frac{4}{3}\,T_F\,n_f\right)\\ 
&+\left(\frac{a_s}{\pi}\right)^2
\left(C_A^2 \left(\ln \left(\frac{\mu }{Q}\right)\left(\frac{\pi^2}{3}
-\frac{67}{9}\right)+\frac{17 \pi^2}{72}-\frac{839}{108}\right)\right.\\ 
&\left.+C_A\,T_F\,n_f\left(\frac{20}{9} \ln \left(\frac{\mu }{Q}\right)-
\frac{\pi^2}{18}+\frac{79}{27}\right)+C_F\,T_F\,n_f\right) \; .
\end{split}
\end{equation} 
Notice that the anomalous dimension for the adjoint representation of
the initial state, i.e.\ for gluon fusion, is the same for the
symmetric and anti-symmetric final state color octet configuration.

Additionally, we reproduce the necessary soft limits of the splitting
functions in Mellin space, which are, up to this order,
textbook material
\begin{equation}
\begin{split}
P_{{qq}}(N) &=\frac{\alpha_{{s}}}{2\pi}\,C_{{F}} \left(-2
\ln (N)+\frac{3}{2}\right)+\left(\frac{\alpha_{{s}}}{2\pi}\right)^2
\left(C_{{A}} C_{{F}} \left(\left(\frac{\pi ^2}{3}-\frac{67}{9}\right)
\ln (N)-3 \zeta (3)+\frac{11 \pi^2}{18}+\frac{17}{24}\right)\right.\\ 
&\left.+\,C_F^2 \left(6 \zeta(3)+\frac{3}{8}-\frac{\pi ^2}{2}\right)+
C_F\,T_F\,n_f\left(\frac{20}{9} \ln(N)-\frac{2 \pi^2}{9}-
\frac{1}{6}\right)\right) \; ,\\ 
P_{gg}(N)
&=\frac{\alpha_{{s}}}{2\pi}\,\left(C_A \left(-2\ln(N)+\frac{11}{6}\right)-
\frac{2}{3}\,T_F\,n_f\right)+\left(\frac{\alpha_{{s}}}{2\pi}\right)^2
\left(C_A^2 \left(\left(\frac{\pi ^2}{3}-\frac{67}{9}\right) \ln(N)+3
\zeta(3)+\frac{8}{3}\right)\right.\\ &\left.+
\,C_A\,T_F\,n_f\,\left(\frac{20}{9}\ln(N)-
\frac{4}{3}\right)-C_F\,T_F\,n_f\right) 
\; ,
\end{split}
\end{equation}
where $f(N) = \int_0^1 dz \; z^{\tilde{N}-1}f(z)$ and
$\tilde N = N e^{-\gamma_E}$. As is well known, the off-diagonal terms
of the splitting functions are not singular in the soft limit.

%%%%%%%%%%%%%%%%%%%%%%%%%%%%%%%%%%%%%%%%%%%%%%%%%%%%%%%%%%%%%%%%%%%%%%%%%%%%%%%%


\section*{References}

\begin{thebibliography}{00}

%\cite{Baernreuther:2012ws}
\bibitem{Baernreuther:2012ws}
  P.~B\"arnreuther, M.~Czakon and A.~Mitov,
  %``Percent Level Precision Physics at the Tevatron: First Genuine NNLO QCD Corrections to $q \bar{q} \to t \bar{t} + X$,''
  Phys.\ Rev.\ Lett.\  {\bf 109} (2012) 132001
  [arXiv:1204.5201 [hep-ph]].
  %%CITATION = ARXIV:1204.5201;%%

%\cite{Czakon:2012zr}
\bibitem{Czakon:2012zr}
  M.~Czakon and A.~Mitov,
  %``NNLO corrections to top-pair production at hadron colliders: the all-fermionic scattering channels,''
  JHEP {\bf 1212} (2012) 054
  [arXiv:1207.0236 [hep-ph]].
  %%CITATION = ARXIV:1207.0236;%%

%\cite{Czakon:2012pz}
\bibitem{Czakon:2012pz}
  M.~Czakon and A.~Mitov,
  %``NNLO corrections to top pair production at hadron colliders: the quark-gluon reaction,''
  JHEP {\bf 1301} (2013) 080
  [arXiv:1210.6832 [hep-ph]].
  %%CITATION = ARXIV:1210.6832;%%

%\cite{Czakon:2013goa}
\bibitem{Czakon:2013goa}
  M.~Czakon, P.~Fiedler and A.~Mitov,
  %``The total top quark pair production cross-section at hadron colliders through O(alpha_S^4),''
  Phys.\ Rev.\ Lett.\  {\bf 110} (2013) 252004
  [arXiv:1303.6254 [hep-ph]].
  %%CITATION = ARXIV:1303.6254;%%

%\cite{Beneke:2009ye}
\bibitem{Beneke:2009ye}
  M.~Beneke, M.~Czakon, P.~Falgari, A.~Mitov and C.~Schwinn,
  %``Threshold expansion of the gg(qq-bar) ---> QQ-bar + X cross section at O(alpha(s)**4),''
  Phys.\ Lett.\ B {\bf 690} (2010) 483
  [arXiv:0911.5166 [hep-ph]].
  %%CITATION = ARXIV:0911.5166;%%

%\cite{Czakon:2009zw}
\bibitem{Czakon:2009zw}
  M.~Czakon, A.~Mitov and G.~F.~Sterman,
  %``Threshold Resummation for Top-Pair Hadroproduction to Next-to-Next-to-Leading Log,''
  Phys.\ Rev.\ D {\bf 80} (2009) 074017
  [arXiv:0907.1790 [hep-ph]].
  %%CITATION = ARXIV:0907.1790;%%

%\cite{Beneke:2009rj}
\bibitem{Beneke:2009rj}
  M.~Beneke, P.~Falgari and C.~Schwinn,
  %``Soft radiation in heavy-particle pair production: All-order colour structure and two-loop anomalous dimension,''
  Nucl.\ Phys.\ B {\bf 828} (2010) 69
  [arXiv:0907.1443 [hep-ph]].
  %%CITATION = ARXIV:0907.1443;%%

%\cite{Beneke:2010da}
\bibitem{Beneke:2010da}
  M.~Beneke, P.~Falgari and C.~Schwinn,
  %``Threshold resummation for pair production of coloured heavy (s)particles at hadron colliders,''
  Nucl.\ Phys.\ B {\bf 842} (2011) 414
  [arXiv:1007.5414 [hep-ph]].
  %%CITATION = ARXIV:1007.5414;%%

%\cite{Cacciari:2011hy}
\bibitem{Cacciari:2011hy}
  M.~Cacciari, M.~Czakon, M.~Mangano, A.~Mitov and P.~Nason,
  %``Top-pair production at hadron colliders with next-to-next-to-leading logarithmic soft-gluon resummation,''
  Phys.\ Lett.\ B {\bf 710} (2012) 612
  [arXiv:1111.5869 [hep-ph]].
  %%CITATION = ARXIV:1111.5869;%%

%\cite{Beneke:2011mq}
\bibitem{Beneke:2011mq}
  M.~Beneke, P.~Falgari, S.~Klein and C.~Schwinn,
  %``Hadronic top-quark pair production with NNLL threshold resummation,''
  Nucl.\ Phys.\ B {\bf 855} (2012) 695
  [arXiv:1109.1536 [hep-ph]].
  %%CITATION = ARXIV:1109.1536;%%

%\cite{Beneke:2012wb}
\bibitem{Beneke:2012wb}
  M.~Beneke, P.~Falgari, S.~Klein, J.~Piclum, C.~Schwinn, M.~Ubiali and F.~Yan,
  %``Inclusive Top-Pair Production Phenomenology with TOPIXS,''
  JHEP {\bf 1207} (2012) 194
  [arXiv:1206.2454 [hep-ph]].
  %%CITATION = ARXIV:1206.2454;%%

%\cite{Czakon:2008ii}
\bibitem{Czakon:2008ii}
  M.~Czakon and A.~Mitov,
  %``Inclusive Heavy Flavor Hadroproduction in NLO QCD: The Exact Analytic Result,''
  Nucl.\ Phys.\ B {\bf 824} (2010) 111
  [arXiv:0811.4119 [hep-ph]].
  %%CITATION = ARXIV:0811.4119;%%

%\cite{Czakon:2008cx}
\bibitem{Czakon:2008cx}
  M.~Czakon and A.~Mitov,
  %``On the Soft-Gluon Resummation in Top Quark Pair Production at Hadron Colliders,''
  Phys.\ Lett.\ B {\bf 680} (2009) 154
  [arXiv:0812.0353 [hep-ph]].
  %%CITATION = ARXIV:0812.0353;%%

%\cite{Belitsky:1998tc}
\bibitem{Belitsky:1998tc}
  A.~V.~Belitsky,
  %``Two loop renormalization of Wilson loop for Drell-Yan production,''
  Phys.\ Lett.\ B {\bf 442} (1998) 307
  [hep-ph/9808389].
  %%CITATION = HEP-PH/9808389;%%

%\cite{Idilbi:2009cc}
\bibitem{Idilbi:2009cc}
  A.~Idilbi, C.~Kim and T.~Mehen,
  %``Factorization and resummation for single color-octet scalar production at the LHC,''
  Phys.\ Rev.\ D {\bf 79} (2009) 114016
  [arXiv:0903.3668 [hep-ph]].
  %%CITATION = ARXIV:0903.3668;%%

%\cite{Ferroglia:2012ku}
\bibitem{Ferroglia:2012ku}
  A.~Ferroglia, B.~D.~Pecjak and L.~L.~Yang,
  %``Soft-gluon resummation for boosted top-quark production at hadron colliders,''
  Phys.\ Rev.\ D {\bf 86} (2012) 034010
  [arXiv:1205.3662 [hep-ph]].
  %%CITATION = ARXIV:1205.3662;%%

%\cite{Ferroglia:2012uy}
\bibitem{Ferroglia:2012uy}
  A.~Ferroglia, B.~D.~Pecjak, L.~L.~Yang, B.~D.~Pecjak and L.~L.~Yang,
  %``The NNLO soft function for the pair invariant mass distribution of boosted top quarks,''
  JHEP {\bf 1210} (2012) 180
  [arXiv:1207.4798 [hep-ph]].
  %%CITATION = ARXIV:1207.4798;%%

%\cite{Ferroglia:2013zwa}
\bibitem{Ferroglia:2013zwa} 
  A.~Ferroglia, B.~D.~Pecjak and L.~L.~Yang,
  %``Top-quark pair production at high invariant mass: an NNLO soft plus virtual approximation,''
  JHEP {\bf 1309}, 032 (2013)
  [arXiv:1306.1537 [hep-ph]].
  %%CITATION = ARXIV:1306.1537;%%

%\cite{Ferroglia:2013awa}
\bibitem{Ferroglia:2013awa}
  A.~Ferroglia, S.~Marzani, B.~D.~Pecjak and L.~L.~Yang,
  %``Boosted top production: factorization and resummation for single-particle inclusive distributions,''
  arXiv:1310.3836 [hep-ph].
  %%CITATION = ARXIV:1310.3836;%%

%\cite{Zhu:2012ts}
\bibitem{Zhu:2012ts}
  H.~X.~Zhu, C.~S.~Li, H.~T.~Li, D.~Y.~Shao and L.~L.~Yang,
  %``Transverse-momentum resummation for top-quark pairs at hadron colliders,''
  Phys.\ Rev.\ Lett.\  {\bf 110} (2013) 082001
  [arXiv:1208.5774 [hep-ph]].
  %%CITATION = ARXIV:1208.5774;%%

%\cite{Li:2013mia}
\bibitem{Li:2013mia}
  H.~T.~Li, C.~S.~Li, D.~Y.~Shao, L.~L.~Yang and H.~X.~Zhu,
  %``Top quark pair production at small transverse momentum in hadronic collisions,''
  Phys.\ Rev.\ D {\bf 88} (2013) 074004
  [arXiv:1307.2464 [hep-ph]].
  %%CITATION = ARXIV:1307.2464;%%

%\cite{Ahrens:2011mw}
\bibitem{Ahrens:2011mw}
  V.~Ahrens, A.~Ferroglia, M.~Neubert, B.~D.~Pecjak and L.~-L.~Yang,
  %``RG-improved single-particle inclusive cross sections and forward-backward asymmetry in $t\bar t$ production at hadron colliders,''
  JHEP {\bf 1109} (2011) 070
  [arXiv:1103.0550 [hep-ph]].
  %%CITATION = ARXIV:1103.0550;%%

%\cite{Kidonakis:2010dk}
\bibitem{Kidonakis:2010dk}
  N.~Kidonakis,
  %``Next-to-next-to-leading soft-gluon corrections for the top quark cross section and transverse momentum distribution,''
  Phys.\ Rev.\ D {\bf 82} (2010) 114030
  [arXiv:1009.4935 [hep-ph]].
  %%CITATION = ARXIV:1009.4935;%%

%\cite{Ferroglia:2009ii}
\bibitem{Ferroglia:2009ii}
  A.~Ferroglia, M.~Neubert, B.~D.~Pecjak and L.~L.~Yang,
  %``Two-loop divergences of massive scattering amplitudes in non-abelian gauge theories,''
  JHEP {\bf 0911} (2009) 062
  [arXiv:0908.3676 [hep-ph]].
  %%CITATION = ARXIV:0908.3676;%%

%\cite{Aybat:2006mz}
\bibitem{Aybat:2006mz}
  S.~M.~Aybat, L.~J.~Dixon and G.~F.~Sterman,
  %``The Two-loop soft anomalous dimension matrix and resummation at next-to-next-to leading pole,''
  Phys.\ Rev.\ D {\bf 74} (2006) 074004
  [hep-ph/0607309].
  %%CITATION = HEP-PH/0607309;%%

%\cite{Mitov:2009sv}
\bibitem{Mitov:2009sv}
  A.~Mitov, G.~F.~Sterman and I.~Sung,
  %``The Massive Soft Anomalous Dimension Matrix at Two Loops,''
  Phys.\ Rev.\ D {\bf 79} (2009) 094015
  [arXiv:0903.3241 [hep-ph]].
  %%CITATION = ARXIV:0903.3241;%%

%\cite{Becher:2009kw}
\bibitem{Becher:2009kw}
  T.~Becher and M.~Neubert,
  %``Infrared singularities of QCD amplitudes with massive partons,''
  Phys.\ Rev.\ D {\bf 79} (2009) 125004
   [Erratum-ibid.\ D {\bf 80} (2009) 109901]
  [arXiv:0904.1021 [hep-ph]].
  %%CITATION = ARXIV:0904.1021;%%

%\cite{Mitov:2010xw}
\bibitem{Mitov:2010xw}
  A.~Mitov, G.~F.~Sterman and I.~Sung,
  %``Computation of the Soft Anomalous Dimension Matrix in Coordinate Space,''
  Phys.\ Rev.\ D {\bf 82} (2010) 034020
  [arXiv:1005.4646 [hep-ph]].
  %%CITATION = ARXIV:1005.4646;%%

%\cite{Cvitanovic:1976am}
\bibitem{Cvitanovic:1976am}
  P.~Cvitanovic,
  %``Group theory for Feynman diagrams in non-Abelian gauge theories,''
  Phys.\ Rev.\ D {\bf 14} (1976) 1536.
  %%CITATION = PHRVA,D14,1536;%%

%\cite{Mitov:2010rp}
\bibitem{Mitov:2010rp}
  A.~Mitov, G.~Sterman and I.~Sung,
  %``Diagrammatic Exponentiation for Products of Wilson Lines,''
  Phys.\ Rev.\ D {\bf 82} (2010) 096010
  [arXiv:1008.0099 [hep-ph]].
  %%CITATION = ARXIV:1008.0099;%%

%\cite{Gardi:2013ita}
\bibitem{Gardi:2013ita}
  E.~Gardi, J.~M.~Smillie and C.~D.~White,
  %``The Non-Abelian Exponentiation theorem for multiple Wilson lines,''
  JHEP {\bf 1306} (2013) 088
  [arXiv:1304.7040 [hep-ph]].
  %%CITATION = ARXIV:1304.7040;%%

\bibitem{Somogyi:2011ir}
  G.~Somogyi,
  %``Angular integrals in d dimensions,''
  J.\ Math.\ Phys.\  {\bf 52} (2011) 083501
  [arXiv:1101.3557 [hep-ph]].
  %%CITATION = ARXIV:1101.3557;%%

%\cite{Czakon:2005rk}
\bibitem{Czakon:2005rk}
  M.~Czakon,
  %``Automatized analytic continuation of Mellin-Barnes integrals,''
  Comput.\ Phys.\ Commun.\  {\bf 175} (2006) 559
  [hep-ph/0511200].
  %%CITATION = HEP-PH/0511200;%%

%\cite{Huber:2005yg}
\bibitem{Huber:2005yg}
  T.~Huber and D.~Maitre,
  %``HypExp: A Mathematica package for expanding hypergeometric functions around integer-valued parameters,''
  Comput.\ Phys.\ Commun.\  {\bf 175} (2006) 122
  [hep-ph/0507094].
  %%CITATION = HEP-PH/0507094;%%

%\cite{Bierenbaum:2011gg}
\bibitem{Bierenbaum:2011gg}
  I.~Bierenbaum, M.~Czakon and A.~Mitov,
  %``The singular behavior of one-loop massive QCD amplitudes with one external soft gluon,''
  Nucl.\ Phys.\ B {\bf 856} (2012) 228
  [arXiv:1107.4384 [hep-ph]].
  %%CITATION = ARXIV:1107.4384;%%

\end{thebibliography}
\end{document}